\documentclass{article}

\usepackage[svgnames, x11names]{xcolor}

\newcommand{\ysnoted}[1]{}

\newcommand{\note}[1]{ {\textcolor{red} {#1}}}

\usepackage[normalem]{ulem} 

\usepackage{listings}

\lstdefinelanguage{XML}
{
basicstyle=\ttfamily\footnotesize,
  morestring=[b]",
  moredelim=[s][\bfseries\color{Maroon}]{<}{\ },
  moredelim=[s][\bfseries\color{Maroon}]{</}{>},
  moredelim=[l][\bfseries\color{Maroon}]{/>},
  moredelim=[l][\bfseries\color{Maroon}]{>},
  morecomment=[s]{<?}{?>},
  morecomment=[s]{<!--}{-->},
  commentstyle=\color{gray},
  stringstyle=\color{blue},
  identifierstyle=\color{red}
}
\usepackage[pdftex]{graphicx}
\graphicspath{{./figures/}}
\DeclareGraphicsExtensions{.pdf}

\usepackage[cmex10]{amsmath}
\usepackage{amssymb}
\usepackage{mathtools}

\usepackage[]{subfig}

\usepackage{algorithmicx}
\usepackage{algpseudocode}
\usepackage[ruled]{algorithm}
\definecolor{light-gray}{gray}{0.75}
\algrenewcommand{\algorithmiccomment}[1]{\hskip3em{{\footnotesize \textcolor{light-gray}{$\blacktriangleright$}}} #1}

\usepackage{multirow}

\usepackage{hyperref}

\usepackage{xspace}
\usepackage[nocompress]{cite}

\usepackage{enumitem}

\usepackage[english]{babel}
\usepackage{blindtext}

\begin{document}
%
\title{\emph{ARM Wrestling with Big Data:}\\A Study of Commodity ARM64 Server\\for Big Data Workloads\footnote{Accepted for publication in the Proceedings of the \emph{24th IEEE International Conference on High Performance Computing, Data, and Analytics (HiPC)}, 2017}}
%
%
%
%

\author{Jayanth Kalyanasundaram 
and Yogesh Simmhan\\
\small
\emph{Department of Computational and Data Sciences,}\\ 
\small \emph{Indian Institute of Science, Bangalore 560012, India}\\
\small \emph{Email: jayantkalyanasundaram@gmail.com, 
simmhan@cds.iisc.ac.in}
}

\maketitle

\begin{abstract}
ARM processors have dominated the mobile device market in the last decade due to their favorable computing to energy ratio. In this age of Cloud data centers and Big Data analytics, the focus is increasingly on power efficient processing, rather than just high throughput computing. ARM's first commodity server-grade processor is the recent \emph{AMD A1100}-series processor, based on a 64-bit ARM Cortex A57 architecture. In this paper, we study the performance and energy efficiency of a server based on this ARM64 CPU, relative to a comparable server running an \emph{AMD Opteron 3300}-series x64 CPU, for Big Data workloads. Specifically, we study these for Intel's HiBench suite of web, query and machine learning benchmarks on Apache Hadoop v2.7 in a pseudo-distributed setup, for data sizes up to $20GB$ files, $5M$ web pages and $500M$ tuples. Our results show that the ARM64 server's runtime performance is comparable to the x64 server for integer-based workloads like Sort and Hive queries, and only lags behind for floating-point intensive benchmarks like PageRank, when they do not exploit data parallelism adequately. We also see that the ARM64 server takes $\frac{1}{3}^{rd}$ the energy, and has an Energy Delay Product (EDP) that is $50-71\%$ lower than the x64 server. These results hold promise for ARM64 data centers hosting Big Data workloads to reduce their operational costs, while opening up opportunities for further analysis.
\end{abstract}


\section{Introduction} 
Mobile and Cloud computing have transformed computing in the 21$^{st}$ century, with millions of servers currently hosted at public data centers and billions of smart phones in the hands of consumers. At the same time, these two classes of computing devices have been supported by two different categories of processor architectures. \emph{x64 processors (also called x86-64)} have traditionally held sway over Cloud servers, with Intel and AMD offering 64-bit versions of their x86 instruction sets based on a CISC architecture, and with support for hardware virtualization. ARM's RISC-based processor architectures, on the other hand, have dominated mobile platforms, including smart phones, tablets, and embedded Internet of Things (IoT) devices, typically running a 32-bit instruction set. 

The RISC architecture natively offers a lower power envelope relative to CISC processors while having more limited performance~\cite{blem2013power}. At the same time, x64 processors have been increasing the number of cores per processor to overcome the power-wall that limits their performance growth in single-core clock speeds. As a result, the performance difference between individual ARM and x64 cores has been narrowing. 

ARM recently introduced their ARMv8-A architecture with a 64-bit instruction set, \emph{ARM64 (also called AArch64)}, to increase the memory addressing available to their processors. In 2016, AMD released the first ARM64 System on Chip (SoC), the A1100 series processor with 8-cores (code named \emph{Seattle}), using ARM's Cortex-A57 micro-architecture~\cite{gwennap2014thunderx}. Since energy consumption by servers forms the major fraction of the operational cost for Cloud data centers, ARM64 with its lower energy footprint and server-grade memory addressing has started to become a viable platform for 
scale-out (rather than scale-up) workloads that are common in Clouds applications, and the growing trend of containerization as opposed to virtualization.

However, given their recent emergence, there has been no literature offering an empirical study of the effectiveness of ARM64 processors for Cloud-based applications, and specifically for Big Data workloads, to confirm these possibilities. While 32-bit ARM processors have been explored for diverse workloads~\cite{Loghin-vldb-2015,rajovic2013mobile}, there has been no study of servers based on the new ARM64 processor for Big Data workloads. Indeed, we could not identify any relevant publication using the AMD A1100 commodity SoC. To our knowledge, the only research study on the performance of an ARM A57 processor is for an HPC workload using an AppliedMicro X-Gene server~\cite{laurenzano:ispass:2016}. 

IaaS Cloud providers are rapidly diversifying into heterogeneous compute offerings such as VMs with GPU accelerators, HPC inter-connects, containerized deployments, and spot-pricing, besides their on-demand VMs hosted on commodity hardware. Given this rapidly evolving eco-system, an energy-efficient ARM64-based compute offering that is competitively priced and performant, is highly conceivable~\cite{rodero2010energy}. Understanding the value of such a platform for end-users with data-intensive workloads is, thus, crucial. We address this gap with an experimental study offering early insights on the computational scaling and energy efficiency an ARM64 server for Big Data workloads. 

Our goal here is \emph{not} to delve into the internal architecture of the ARM64 processor, which is well-studied elsewhere, but rather consider it's impact for a data science end-user, a Big Data platform developer or an Infrastructure as a Service (IaaS) provider. Hence, we are interested in the workload's compute and energy performance outcomes to help users choose between comparable ARM64 and x64 servers.
We make the following specific contributions in this paper.
\begin{enumerate}
\item We introduce the system model for two comparable servers, running ARM64 and x64 CPUs~(\S~\ref{sec:system}), 
and describe Intel's HiBench Big Data benchmark suite~\cite{hibench} used to evaluate these servers 
for diverse workload sizes~(\S~\ref{sec:workloads}).
\item We offer a detailed performance evaluation of these two servers, for text analytics and I/O \emph{micro-benchmarks}, and web, query, and machine learning \emph{application benchmarks}~(\S~\ref{sec:perf}). We present an analysis on these results based on platform and hardware features. 
\item We report the power consumption by the servers for these workloads, and their Energy Delay Product (EDP) that impacts their operational energy efficiency~(\S~\ref{sec:energy}).
\end{enumerate}

\section{Related Work}

32-bit ARM processors (\emph{ARM32}) were designed for mobile and embedded devices, such as smart phones and Raspberry Pis, and have also been used in network switches in the data center. 
Their focus has traditionally been on power efficiency, as 
opposed to server-grade processors that focus on performance. 
There have been several studies of ARM32 processors, and their comparison with x86 or x64 processors, as we discuss below. But none explore an ARM64 commodity server which will be widely available, nor its effectiveness for Big Data workloads, which we address in this paper.


\cite{aroca-jpdc-2012} uses a low cost testing system to systematically compare several ARM and x86 devices. It analyzes the system's power efficiency and CPU performance for a web server, database server and floating point computation. 
The results conclude that ARM is $3-4\times$ more power efficient on a performance per energy comparison although its performance deteriorates when we increase the size of the workload.

Query processing (TPC-C and TPC-H on MySQL) and Big Data (K-Means, TeraSort and WordCount) benchmarks have been tried on the ARM Cortex-A7 little/ARM Cortex-A15 big cores, to compare against Intel Xeon servers~\cite{Loghin-vldb-2015}. They evaluate execution time, energy usage and total cost of running the workloads on these self-hosted ARM and Xeon nodes. 
Their results show that ARM takes more time to perform MapReduce computations 
compared to others, when implemented in Java. However, this time is reduced significantly by the C++ implementation of the same. Our study shows similar limitations with floating-point performance in ARM64 as well.

HPC workloads have been verified for the ARM architecture as well~\cite{rajovic2013mobile}. In \cite{Jarus-lsds-2013}, the performance and energy of HPC clusters 
based on Intel Atom, Core i7 and Xeon E7, AMD G-T40N processors, and ARM Cortex A9 processors was examined. They perform benchmarks like Phoronix Test Suite, CoreMark, Fhourstones, and High-Performance Linpack (HPL).

Analytical models have also been developed to study the performance and energy efficiency of ARM processors. 
\cite{Tudor-sigmetrics-2013}~proposes a model for energy usage of server workloads on low-power multi-core systems like ARM, and validates this for the ARM Cortex-A9 CPU. It uses insights on the ARM architecture to predict the CPU performance analytically. But its evaluation skews toward floating-point workloads on ARM32, and only \emph{memcached} is considered as a Big Data workload.   

Others have offered a concurrency and performance model for ARMv8, with support for 64-bit instructions, to allow verification and analysis over the processor behavior~\cite{flur2016modelling}. Their micro-architectural flow model allows users to investigate concurrency semantics based on an Instruction Set Architecture, but does not deal with an empirical analysis of an actual physical processor. Our goal is to understand the actual performance behavior of a commodity 64-bit ARM server that is likely to be widely used for Big Data workloads and Cloud data centers. One article uses Hidden Markov Models to predict transactional web workloads and plan energy-aware scheduling, validated on the AMD A1100 processor~\cite{ahmed2017dynamic}. Their contribution itself is orthogonal to the ARM64 CPU.


Recently, Laurenzano, \emph{et al.} have studied the ARM64 architecture using the AppliedMicro X-Gene server running the AppliedMicro 883208-X1 CPU~\cite{laurenzano:ispass:2016}. They offer a detailed study of the architecture for different HPC workloads, examining the performance and energy efficiency of the ARM64 platform against Intel's Atom and Xeon architectures. They also offer a model to understand instruction-level behavior that impacts the relative performances. Our study for Big Data workloads offer higher-level insights from the perspective of the application and Hadoop platform rather than instruction-level tuning.  


Many benchmarks for evaluation Big Data applications exist~~\cite{xiong2013characterization}. Big Bench~\cite{ghazal2013bigbench} is a popular benchmark suite that uses an retail eCommerce application as a case study to evaluate processing of high volume and high velocity datasets in the Enterprise. Hibench~\cite{hibench} is another Big Data benchmark from Intel that includes both micro-benchmarks and common application benchmarks from web search, NoSQL queries and Machine Learning. We favor this suite and use it in our evaluation due to the diversity of workloads it targets, for data volume than velocity. There are also benchmarks that specifically target fast data applications, such as for Internet of Things,  
but we defer a study of ARM64 for such Big Data stream processing platforms to future work~\cite{shukla:tpctc:2016}.

BigDataBench compares different Big Data benchmarks including HiBench~\cite{hibench} and BigBench~\cite{ghazal2013bigbench}, and proposes a set of data-intensive applications that is a union of these various existing ones~\cite{wang2014bigdatabench}. They use this to study specific micro-architectural and cache features of the Intel Xeon E5 processor. Our goal is not to analyze the specific internal architecture of ARM. But we do compare the relative performance of the ARM processor on runtime and energy efficiency, compared to an AMD Opteron-based server, for end-users to benefit.

Others~\cite{call:wbdb:2015} have looked at how the cost efficiency of Hadoop jobs on Big Data Platform as a Service (PaaS) can be estimated, using Microsoft Azure's HDInsight as an example. Our work is comparable, and helps estimate the performance and energy costs of running Hadoop workloads on ARM64 servers using an IaaS model. This can in turn inform 
the operational costs for the data center, and plan Cloud pricing for PaaS using ARM64 servers.

\section{Commodity Servers under Test}
\label{sec:system}

The technical specifications of the ARM64 and x64 servers we use in our evaluation are given in Table~\ref{tbl:armrigel}. We try and ensure that the hardware and software configurations of both the systems are identical in most respects, other than for the processor and mother-board distinctions.
\begin{table}[p]
\centering
\caption{Configuration of ARM64 and x64 Nodes} 
\label{tbl:armrigel}
\begin{tabular}{lp{3cm}p{3cm}}
\hline
 & \textbf{ARM64}~$^\dagger$ & \textbf{x64}~$^*$ \\
\hline
\hline
CPU & AMD Opteron A1170 processor, $8\times$ ARM64 A57 cores, 2.0~GHz & AMD Opteron 3380 processor, $8\times$ x64 cores, 2.6~GHz, \emph{Pile Driver}\\
L2/L3 Cache & 4MB/8MB & 8MB/8MB \\
\hline
Memory     & $2 \times 8$~GB DDR3 RAM, Multi Bit ECC   & $2 \times 8$~GB DDR3 RAM, Multi Bit ECC\\
\hline
Disk   &$1TB$ Seagate Barracuda ST1000DM003 HDD, $64MB$ cache (OS \& HDFS)  & $1TB$ Seagate Barracuda ST1000DM003 HDD, $64MB$ cache (HDFS), $256GB$ SSD (OS) \\
\hline
Network & 1 Gbps NIC & 1 Gbps NIC \\
\hline
\hline
Power supply & ATX 200W power supply, No energy rating & 1620W power supply, Platinum level $>90\%$ efficiency \\
Rated TDP & 32W & 65W \\
\hline
\hline
OS       & OpenSUSE Tumbleweed   & CentOS 7 \\
Linux Ver.           & 4.8.6-2-64kb       & 3.10.0-123.20.1.el7.x86\_64\\
\hline
File System	& BTRFS on HDD, $8kb$ block size & EXT4 on SSD, BTRFS on HDD, $4kb$ block size \\ 
Swap      & 512 MB                             & 16 GB\\
\hline
\hline
Java	& OpenJDK 64-bit 1.7.0\_111 (compiled for ARM64)	& OpenJDK 64-bit 1.7.0\_111 (compiled for x64) \\
\hline
Hadoop Ver.	& Apache Hadoop v2.7.3 (pseudo-distributed) (compiled for ARM64)	& Apache Hadoop v2.7.3 (pseudo-distributed) (compiled for x64)\\
HDFS & \multicolumn{2}{c}{$1 \times$ replication factor, 128MB block size}\\
\hline

\end{tabular}
\small{{$^\dagger$~https://shop.softiron.co.uk/product/overdrive-3000/}\\
{$^*$~https://www.supermicro.com/Aplus/system/3U/3012/AS-3012MA-H12TRF.cfm}}
\end{table}

The ARM Cortex-A57 is the first 64-bit server-grade processor design to implement the ARMv8-A architecture, with support for Symmetrical Multiprocessing (SMP) and an out-of-order superscalar pipeline~\footnote{https://developer.arm.com/products/processors/cortex-a/cortex-a57}. AMD's A1100 processor series is the first commercially available commodity chip implementing ARM Cortex-A57, released in Jan, 2016~\footnote{www.amd.com/en-us/products/server/opteron-a-series}. For our evaluation, we use SoftIron's Overdrive 3000 server, which is an Enterprise-class developer system that is based on AMD A1170 processor having 8 cores at 2~GHz, 16~GB RAM, 1~TB Seagate Barracuda HDD with a 64MB cache, and Gigabit Ethernet. The server runs an OpenSUSE Linux distribution and uses BTRFS file system. OpenJDK v7 64-bit ARM edition is natively installed and used by our workloads. The server is supplied power by an SMPS that is rated at 200~Watts.

The x64 server used in our evaluation is a single node in a cluster. It has a similar hardware configuration, except using an AMD Opteron 3380 processor with 8 cores rated at 2.6~GHz. It has 16~GB of RAM, a 256~GB SSD for the OS partition, an identical Seagate 1~TB HDD for HDFS data storage and workload applications, and Gigabit Ethernet. We conduct all our experiments using the 1~TB HDD for HDFS to ensure the disk performance is uniform. The x64 server runs CentOS 7 Linux distribution, EXT4 file system for the SSD, and BTRFS for the HDD. We use the same OpenJDK v7 compiled for x64.

Both servers run Hadoop v2.7.3 in pseudo-distributed mode. For the x64 system, we use the standard Apache Hadoop x64 Release~\footnote{http://hadoop.apache.org/releases.html}. 
For the ARM64, we use the same version of \emph{hadoop-arm64}~\footnote{https://github.com/owlab-exp/hadoop-arm64} which gives binaries compiled for ARM64.

\section{HiBench Big Data Workload}
\label{sec:workloads}
HiBench~\cite{hibench} is a widely-used benchmarking suite from Intel used to evaluate Big Data workloads. We use HiBench with different configurations to evaluate the relative merits of the ARM and x64 servers for data intensive applications. HiBench offers both micro-benchmarks and applications benchmarks from domains such as querying, machine learning and graph processing. While HiBench includes workloads that run on platforms like Hadoop, Spark and Storm, we limit our empirical analysis to the popular Apache Hadoop platform, and perform micro-benchmarks and applications benchmarks on it. As far we can tell, there are no Intel specific optimizations for the benchmarks, making them equally suited for the x64 and ARM64 processors from AMD.

Running a workload in HiBench involves two phases: a \emph{prepare phase} where HiBench generates the necessary input data to run the benchmark, 
and an \emph{execution phase} where the application logic of the benchmark are run over the prepared dataset. We focus our analysis on the execution phase where the actual application logic for the workload is performed. 

Next we briefly describe the HiBench workloads we use, and refer readers to the HiBench~\cite{hibench}\footnote{Intel HiBench Suite, https://github.com/intel-hadoop/HiBench} for more details. 

\subsection{Micro-Benchmarks}
The HiBench micro-benchmarks
are based on common applications packaged with the Hadoop platform. 
\subsubsection{Sort}
This workload sorts a given input file. The Map and Reduce functions in the job use the default \texttt{Identity} Mapper and Reducer classes. The effort itself is spent in the shuffle and sort phase, which this benchmark evaluates. The prepare phase generates files in HDFS that contain key-value string pairs of \emph{varying lengths} using the \texttt{RandomTextWriter} class that ships with Apache Hadoop. 

\subsubsection{TeraSort}
TeraSort is similar to sort, but with two distinctions. First, each random row of input data is exactly $100~bytes$ in length with the first $10~bytes$ of the input used as the sort key. Second, the MapReduce job uses a custom partitioner to ensure that the output of a reduce task $R_i$ is lesser than the output of the reduce task $R_{i+1}$. The input data is prepared using the \texttt{TeraGen} class shipped with Hadoop to generate a specified number of rows. 

\subsubsection{Word Count}
Word Count is a popular MapReduce example, 
and returns the frequency of distinct words in the given input documents. The Map function emits each word in the input as the output key along with its local count, and the Reduce function sums these values based on a \emph{groupBy} operation performed on each unique word during shuffle. The input data is created using the \texttt{RandomTextWriter} class. 

\subsubsection{Enhanced DFSIO (DFSIOe)}
This measures the concurrent read and write throughput of HDFS by multiple Mapper tasks working on independent files. 
The prepare phase generates control files of specified sizes. 
During the benchmark execution, the Hadoop job spawns one Map task for each input file and performs read/write operations on it. The Reduce task then calculates the average I/O throughput.

\subsection{Web Search Benchmarks}
These benchmarks in HiBench are representative of MapReduce applications used in large-scale web search and indexing.

\subsubsection{PageRank}
PageRank is a popular graph centrality algorithm used in web search. 
HiBench uses a MapReduce implementation of PageRank from PEGASUS~\cite{kang2009pegasus},
which has two MapReduce jobs: one that updates the rank values for web pages, and is run repeatedly for a fixed number of iterations $ni=3$, and another, a Map-only job that unrolls and emits the PageRank value for each vertex. This gives a total of $2\times ni + 1$ jobs for this workload. 
The input web data is generated in the prepare phase using a Zipfian distribution, for the keywords from the Linux dictionary word list
, and for the links.

\subsubsection{Nutch Indexing}
This benchmark evaluates the indexing part of a web crawl and search pipeline. It uses the MapReduce implementation of the Indexer from the Apache Nutch project, which runs as a single job. The Map tasks
generate the URL of the site to be indexed as key and its metadata.
The Reduce tasks perform the actual indexing, which includes checking if the page should be indexed, running a scoring function to the page or its linked neighbors, and indexing the keywords and metadata in the page. Its input is similar to PageRank.

\subsection{Hive Query Benchmarks}
Apache Hive is a columnar data warehouse that is built on top of Hadoop. It supports SQL-like OLAP queries that are converted to a Directed Acyclic Graph (DAG) of MapReduce jobs for execution. The query benchmarks 
are defined on two \emph{external tables} in Hive, 
that contain either \emph{rankings} with 
web URLs and their PageRank, 
or \emph{user visits} with 
the source IP, the URLs visited, Ad Revenue generated, visited timestamp, etc., generated using a pseudo-random function. 

\subsubsection{Scan}
The scan benchmark is like a \emph{select} SQL query, and performs a \texttt{'SELECT *'} to 
returns all rows present in the user visits table that are then copied into a new table. 
This translates to 3 MapReduce jobs. 
It measures the performance of Hadoop to copy data in HDFS using Hive queries. 

\subsubsection{Join}
This performs a database \emph{join} between the rankings and the user visits tables, with the web URL field in the former matching the visited URL field in the latter, for all pages in a time range. 
It then performs \emph{group by} and \emph{order by} operations on the source IP and ad revenue. 
These get compiled to $2-3$ MapReduce jobs by Hive, depending on the input size. 

\subsubsection{Aggregation}
This measures the performance of Hive for SQL-style \emph{aggregation} queries over large datasets. 
It determines the sum of the revenues from user visits from each source IP based on the
user visits table. The output is a new table with the source IP and its total revenue. It runs as a single MapReduce job.

\subsection{Machine Learning Benchmarks}
Machine Learning (ML) in an important class of application over Big Data. These benchmarks evaluate Hadoop's performance for two classic knowledge discovery and mining algorithms 
implemented as part of the \emph{Apache Mahout project}~\footnote{http://mahout.apache.org/}. 
\subsubsection{Na\"{i}ve Bayes Classifier Training}
This benchmarks the time to train a Na\"{i}ve Bayes classifier for text documents. 
The training involves \texttt{seq2sparse} which pre-processes the input documents to generate the term frequency-inverse document frequency (tf-idf) vector for each document, and \texttt{trainnb} which trains the Bayes classifier based on these vectorized documents.
The prepare phase generates documents with a zipfian distribution from a dictionary of words in Linux. 
\subsubsection{K-Means Clustering}
K-Means clustering identifies $k$ clusters of closely related items within a multi-dimensional structured dataset. The prepare phase generates samples of item vectors based on a Normal distribution, which are present around $k$ centers that are generated from a uniform distribution, and computes the initial $k$ clusters. 
In the execution phase, the MapReduce job updates the $k$ initial cluster centroids and recalculates the clusters 
iteratively till convergence, or a maximum number of iterations $ni=5$. 
A final Map-only job assigns each input item to a cluster. So a minimum of $2$ MapReduce jobs and maximum of $6$ jobs execute.

\section{Performance Results and Analysis}
\label{sec:perf}
\begin{table}[t]
	\centering
	\caption{Data Size Range for Benchmark Workloads} 
	\label{tbl:benchmarks}
	\begin{tabular}{c c p{4.1cm}}
		\hline
		\textbf{Type} & \textbf{Name} & \textbf{Data size ranges} \\
		\hline
		\hline
		Micro & Sort & $10^5 - 2\times10^{10}~bytes$\\
		Micro & TeraSort & $10^5 - 5\times10^8$ rows, $100~bytes$ each\\
		Micro & Word Count & $10^5 - 2\times10^{10}$ bytes\\
                \hline
		Micro & DFSIOe Read & \multirow{2}{3.95cm}{$16 - 512$ files, $10~MB$ \& $500~MB$ each}\\
		Micro & DFSIOe Write &\\
		\hline
		Web Search & PageRank & $10^4 - 5\times10^6$ web pages\\
		Web Search & Nutch Indexing & ${2.5}\times10^4 - 5\times10^5$ web pages\\
		\hline
		Query & Scan &\multirow{3}{3.95cm}{$10^4 - 5\times10^8$ user visit rows \& ${1.2}\times10^3 - 6\times10^7$ ranking rows}\\
		Query & Join &\\
		Query & Aggregation &\\
		\hline
		ML & Na\"{i}ve Bayes & ${2.5}\times10^4 - 2\times10^5$ web pages\\
		ML & K-Means & $10^4 - 10^8$ samples\\
		\hline
	\end{tabular}
\end{table}

We evaluate the ARM64 and x64 servers 
using the HiBench suite, and report their computational performance in this section. An energy analysis is presented in the next section, \S~\ref{sec:energy}. Further, we also offer an analysis of these results to reason about their behavior, and draw insights on the Hadoop configuration to leverage the behavior of the ARM64 server. 

\begin{figure*}[t!]
	\centering
	\subfloat[Runtime of Sort]{
		\includegraphics[width=0.45\textwidth]{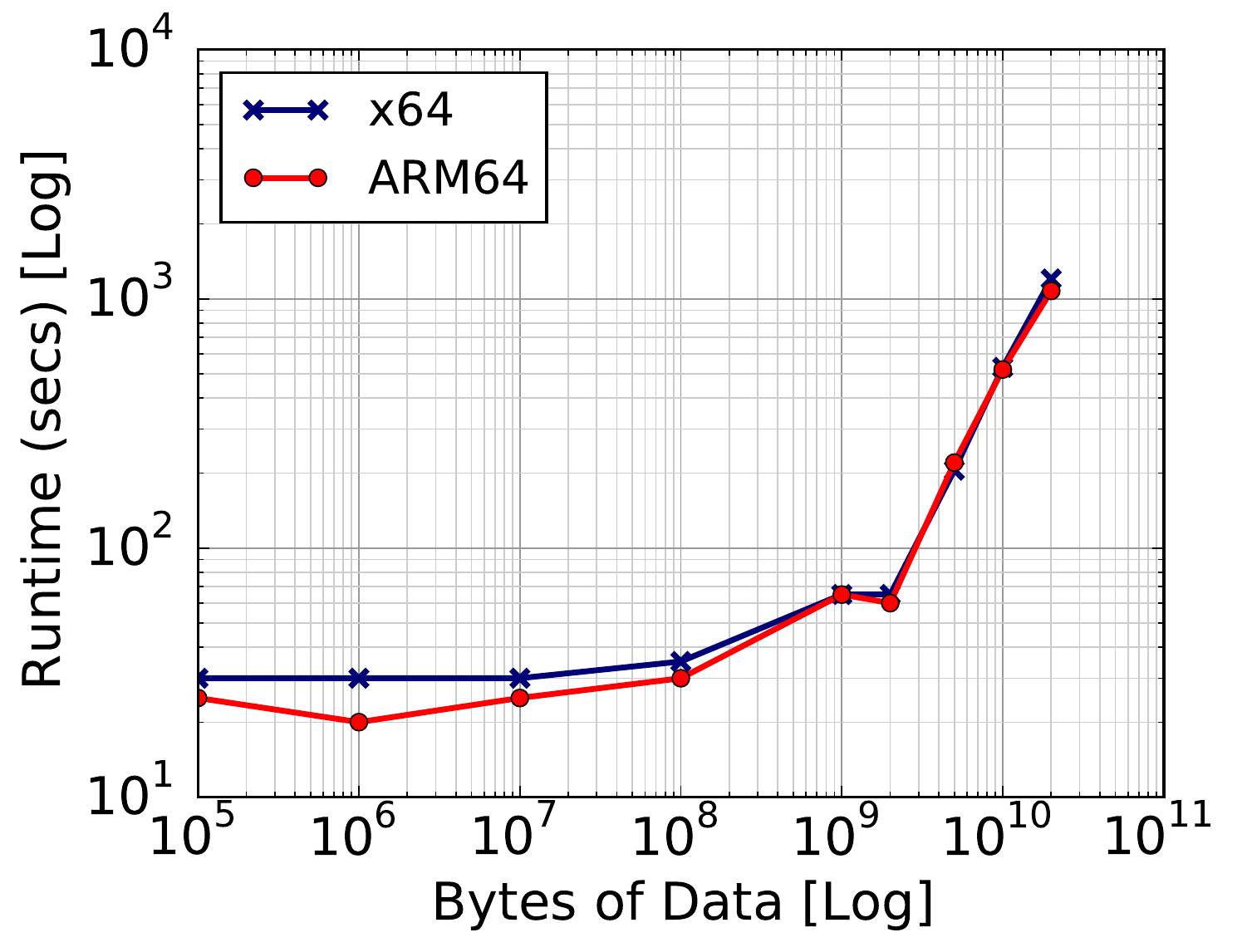} 
		\label{fig:time:sort}		
	}
	\subfloat[Runtime of TeraSort]{
		\includegraphics[width=0.45\textwidth]{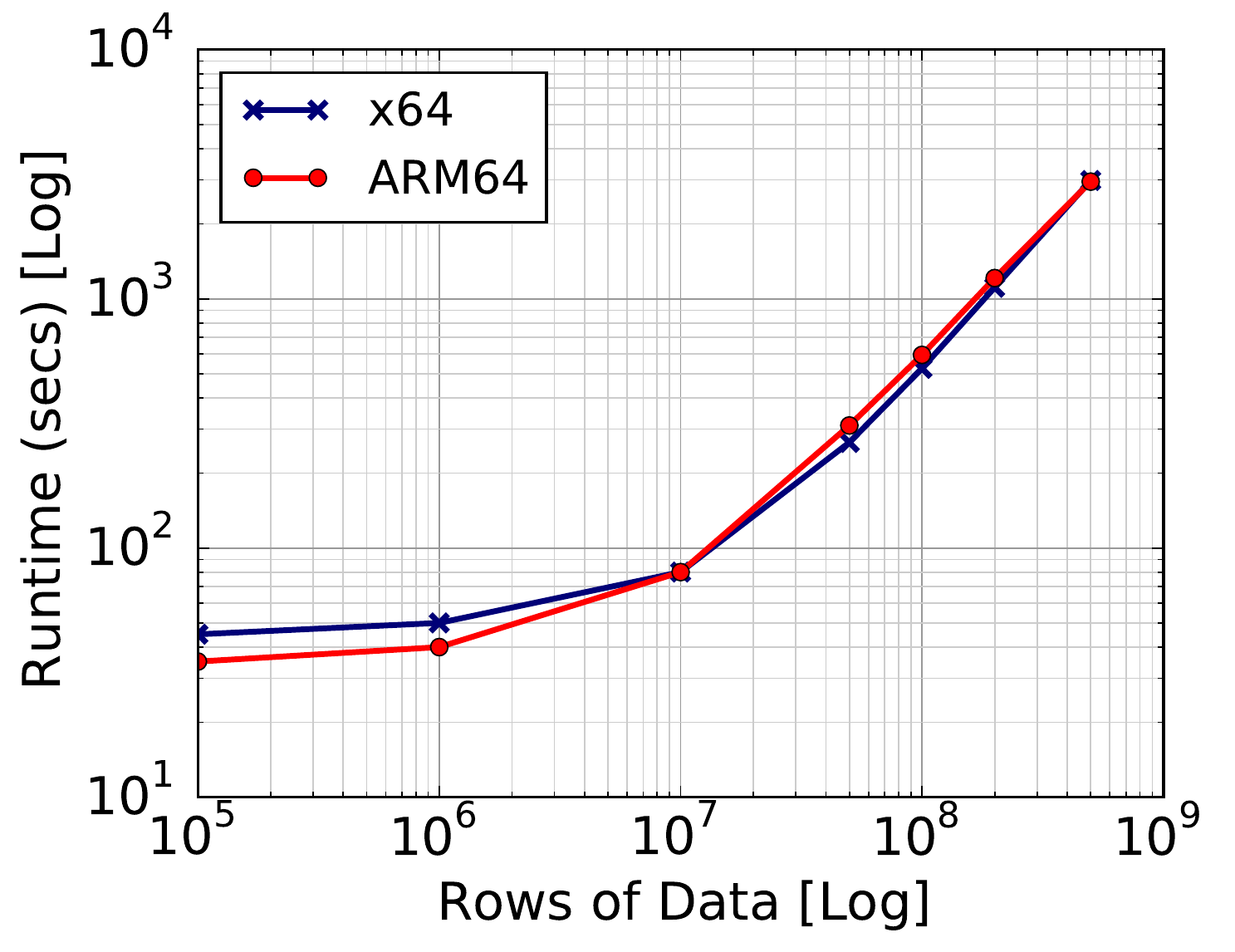} 
		\label{fig:time:tera}
	}\\
	\subfloat[Runtime of Word Count]{
		\includegraphics[width=0.45\textwidth]{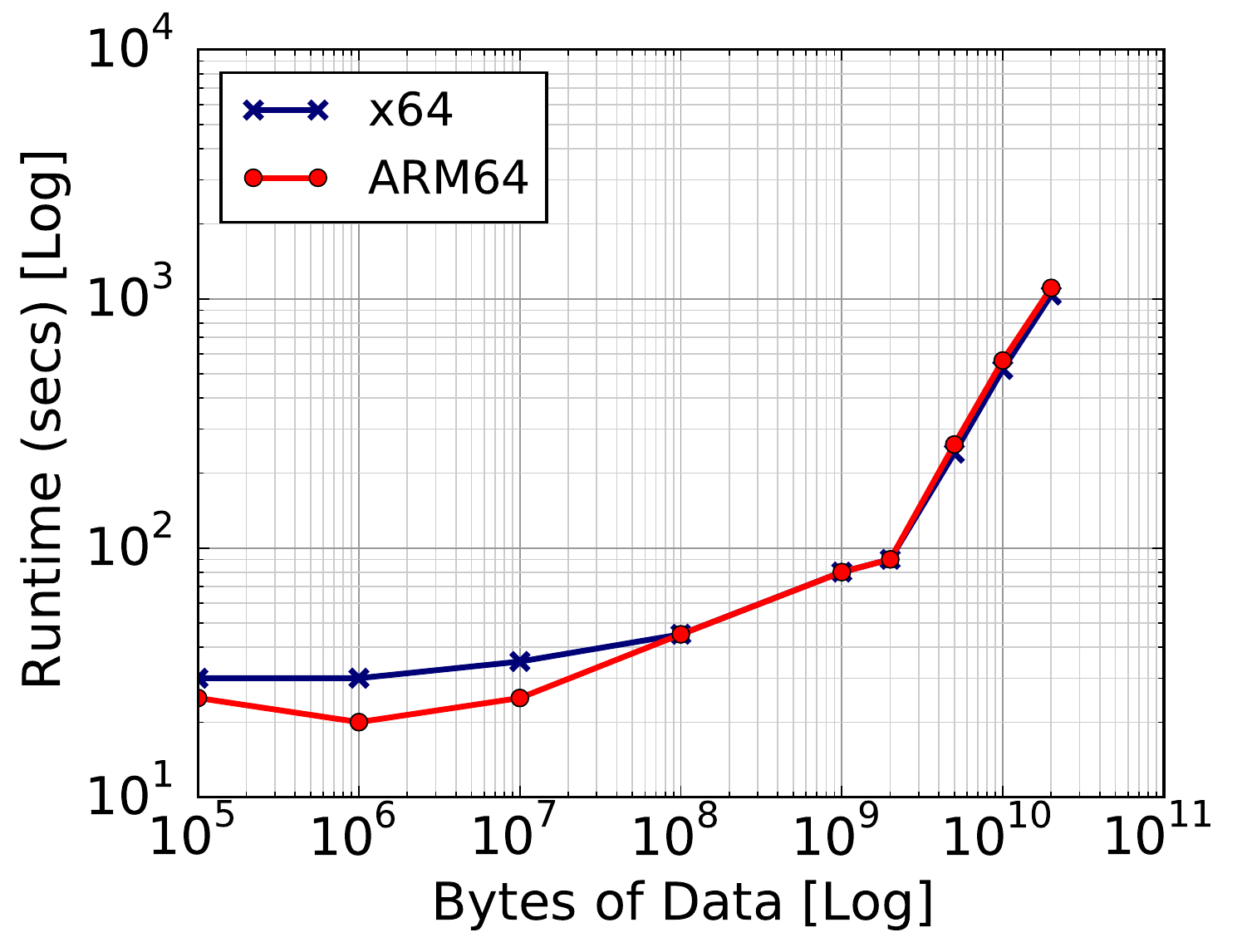}
		\label{fig:time:wc}
	}
	\subfloat[Throughput of DFSIOe Reads \emph{(left)} and Writes \emph{(right)} 
	]{
		\includegraphics[width=0.45\textwidth]{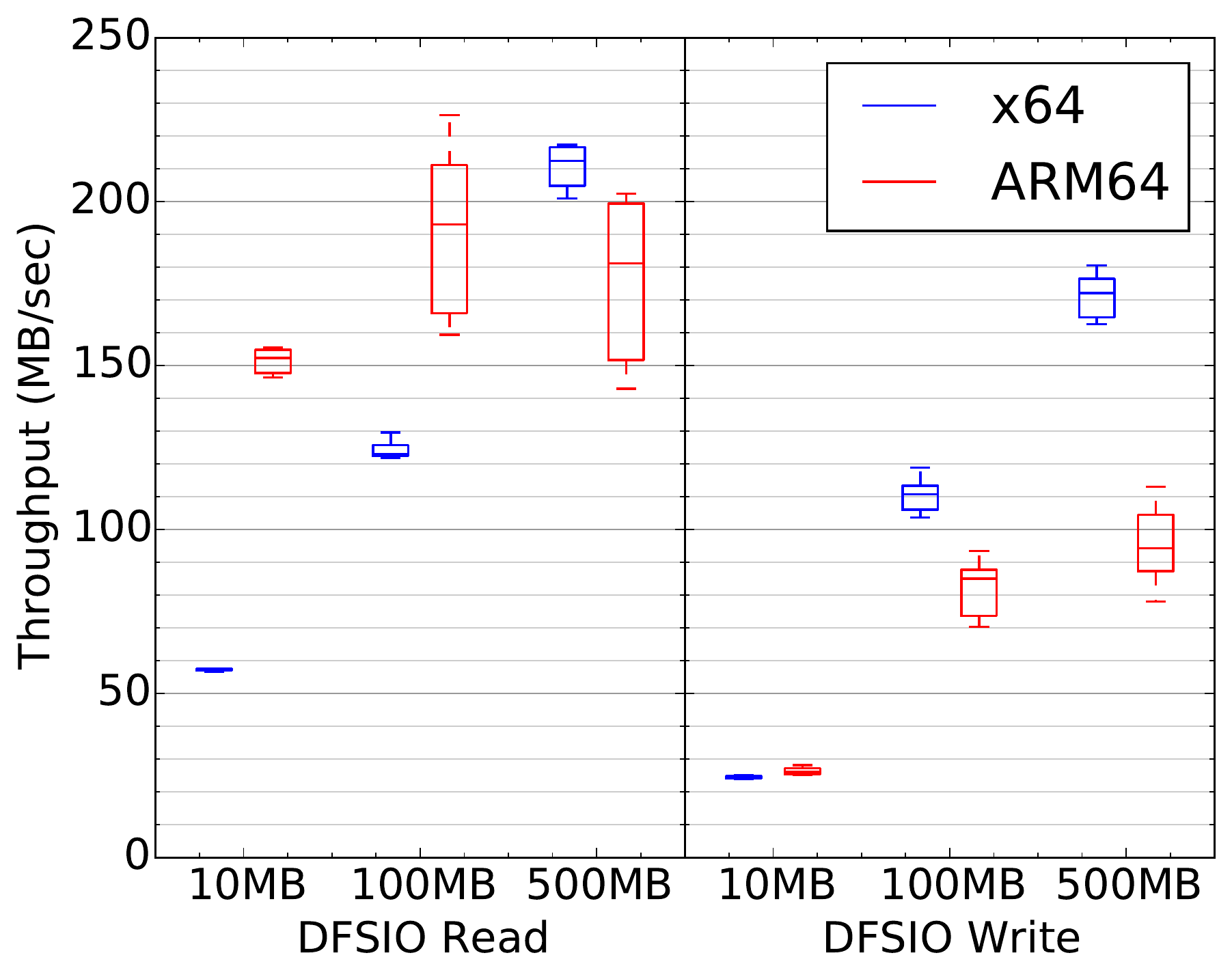}
		\label{fig:thruput:dfsio}
	}
	\caption{Runtime for Sort and Count Micro-Benchmarks, and Throughput for Read and Write from DFSIOe Micro-Benchmarks}
	\label{fig:micro}
\end{figure*}

\subsection{Hadoop Configuration}
We deploy Apache Hadoop v2.7.3 
in a single-node pseudo-distributed setup on the ARM64 and x64 servers. 
HDFS has a replication factor of $1$. 
The resource configuration for the YARN scheduler and MapReduce v2 are based on Enterprise best practices~\cite{hortonconfig}. Specifically, we reserve $1~GB$ for the OS and give YARN access to $15~GB$ of RAM in the nodes. Map/Reduce containers are assigned $1920~MB$/$3840~MB$ of RAM and $1$/$1$ CPU cores, respectively. 
Since the \emph{AppMaster} for a Hadoop job occupies one container, this allows us to run $7$ Map tasks or $3$ Reduce tasks concurrently on a node. We retain the default ratio of $2.1:1$ between the virtual memory and the physical memory for the tasks. 

We found that this configuration was adequate to maximize the data set sizes supported the various benchmarks. However, for the DFSIOe benchmarks which are Map-heavy jobs, the physical memory limit for Map containers
is increased to $3840~MB$ of RAM. 
All the benchmarks are run with an allowed parallelism of $8$ for Map tasks and $4$ for Reduce tasks, but accounting for the AppMaster container, the effective maximum concurrent Map/Reduce tasks is only 7/3.

\subsection{Workload Setup}

We use the publicly available HiBench v5.0 benchmark suite for our experiments. 
Table~\ref{tbl:benchmarks} gives the range of data set sizes used for each benchmark workload, and we exponentially increment the data sizes within this range so that a wide range of values are evaluated. These parameters are defined in the HiBench configuration, and used by its prepare phase to generate required data sizes that are used by the benchmarks.

We use default HiBench values for most benchmark-specific configuration parameters. 
For Terasort,  
we do not compress the Map output to shuffle 
to avoid the computing overhead for compression since we have a single-node setup without network data transfer. 
In addition, we use $10~MB$, $100~MB$ and $500~MB$ files for reads and writes in DFSIOe which are representative of the small, medium and large scales of single-files typically seen in big data workloads.

In the Hive query benchmarks, we maintain a 100:12 ratio between the rows in \emph{user visits} and rows in \emph{web page URLs} table, that matches the ratios for the default data sizes in HiBench. 
We retain as default the number of classes for the Bayes training data as 100. 
The input samples generated for K-Means Clustering 
were defaulted to a 20-dimensional vector. 
The number of clusters $k$ is set to 10 for all sample sizes~\cite{hibench}. 

\ysnoted{Are the map and reduce task time ratios, and execution time ratios similar to HiBench Table 4? How do our times compare against the different stages for a job?}
\ysnoted{network utilization is not a major factor for HiBench. So even for us, the cluster setup can be comparable to single node?}
\ysnoted{Check if the CPU/disk/memory patterns listed for BM in hibench are consistent here.}
\ysnoted{Are we also comparing thruput (jobs/minute) in addition to speed/runtime, like hibench Sec 4.C? How does their old/new processor compare with our opteron/ARM? Intel x5400 vs x5500, harpertown vs nelham gainestown} 

\subsection{Micro-Benchmarks}
Figs.~\ref{fig:time:sort}, \ref{fig:time:tera} and \ref{fig:time:wc} show the times taken (Y Axis, in seconds) for running the sorting and counting micro-benchmarks for different dataset sizes (X Axis). Both axes are in log scale.

For \emph{Sort} and \emph{TeraSort}, we see from Figs.~\ref{fig:time:sort} and \ref{fig:time:tera} that until a data size of $1~GB$ ($10^9~bytes$ or $10^7$ rows), the runtimes are almost flat, and these indicate the data size range where the static overhead of the Hadoop framework out-strips any parallel scaling. 
Beyond this, we see that time taken increases proportional to the data size for both ARM64 and x64 servers.

Between the two servers, we see that ARM is marginally faster for smaller data sizes where the systems are not resource-bound, and beyond that, both the servers perform comparably. 
This trend is sustained for \emph{Word Count} in Fig.~\ref{fig:time:wc}, with the ARM64 server taking lesser time to complete the count for data sizes below $10^8~bytes$ while x64 is up to $6\%$ faster for sizes $\ge5~GB$. Word Count is much more CPU bound than the sorting benchmarks, with both servers using an average $80\%$ CPU for the largest sizes, and performing negligible disk I/O.

\ysnoted{DFSIO does not seem reliable. Replace with our own caculation of thruput. Complement with \texttt{dd}?}
\emph{DFSIOe} performs reads and writes on small, medium and large file sizes, for different numbers of files. Fig.~\ref{fig:thruput:dfsio} shows the box plots of average throughput measured by HiBench from the concurrent Map tasks that perform these reads and writes over independent files in HDFS. 
We see 
that the ARM64 server has a faster cumulative read throughput than the x64, performing almost twice as fast for small and medium file sizes, but marginally slower for $500~MB$ files. While the throughput of ARM64 is less affected by the file size for reads, x64's throughput is proportional to the data size, both for reads and writes. ARM64's write rate is proportional to the file size, but marginally slower than x64. 
Given that the disk specifications and file system are the same for both servers, this behavior may reflect the impact of block sizes or I/O libraries in the OS. 
\ysnoted{Block sizes on ARM is 8k and on x64 is 4k. Does that matter?}

\subsection{Web Search Benchmarks}
\begin{figure}[t!]
	\centering
	\subfloat[PageRank]{
		\includegraphics[width=0.45\columnwidth]{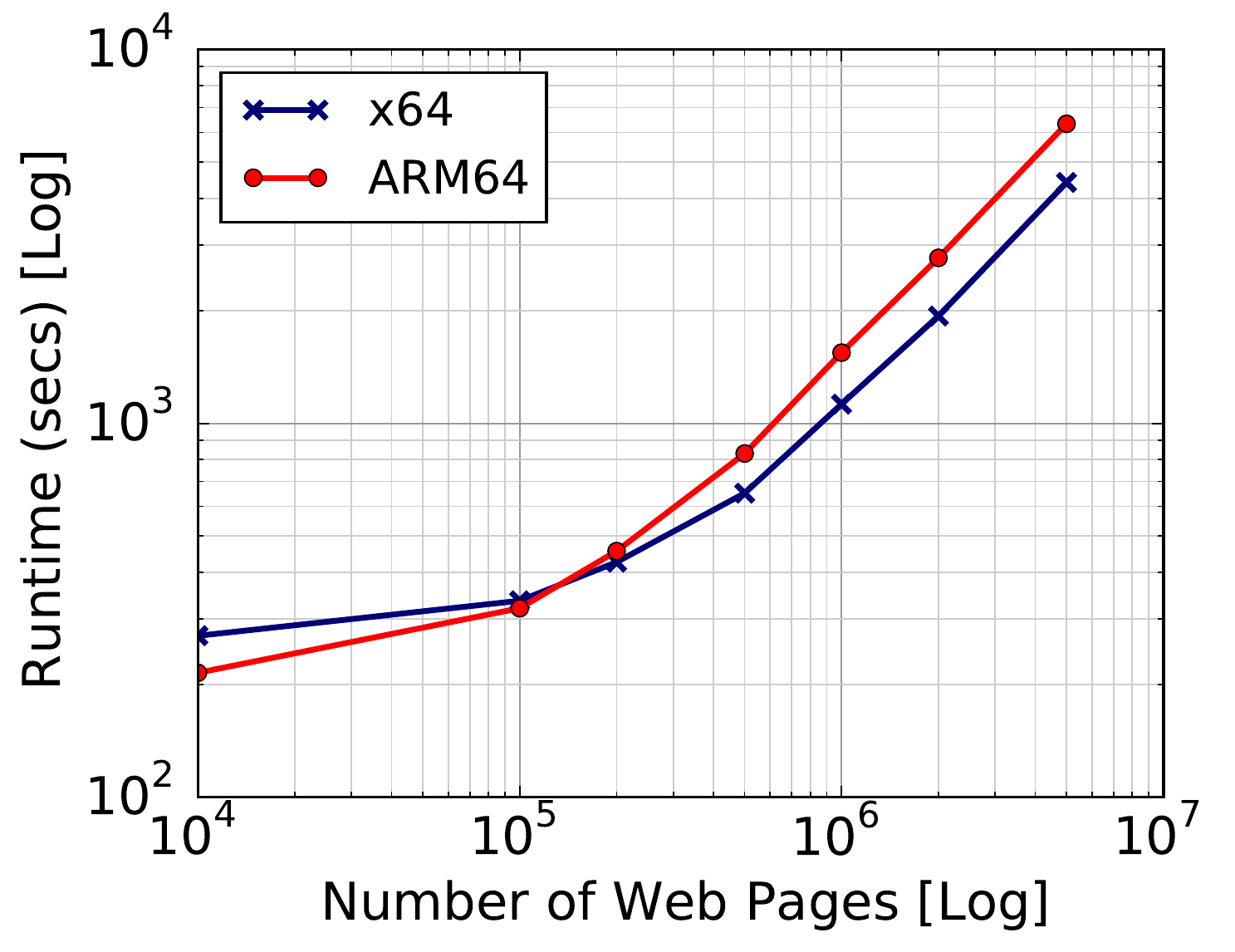}
		\label{fig:time:pr}
	}
	\subfloat[Nutch Indexing]{
		\includegraphics[width=0.45\columnwidth]{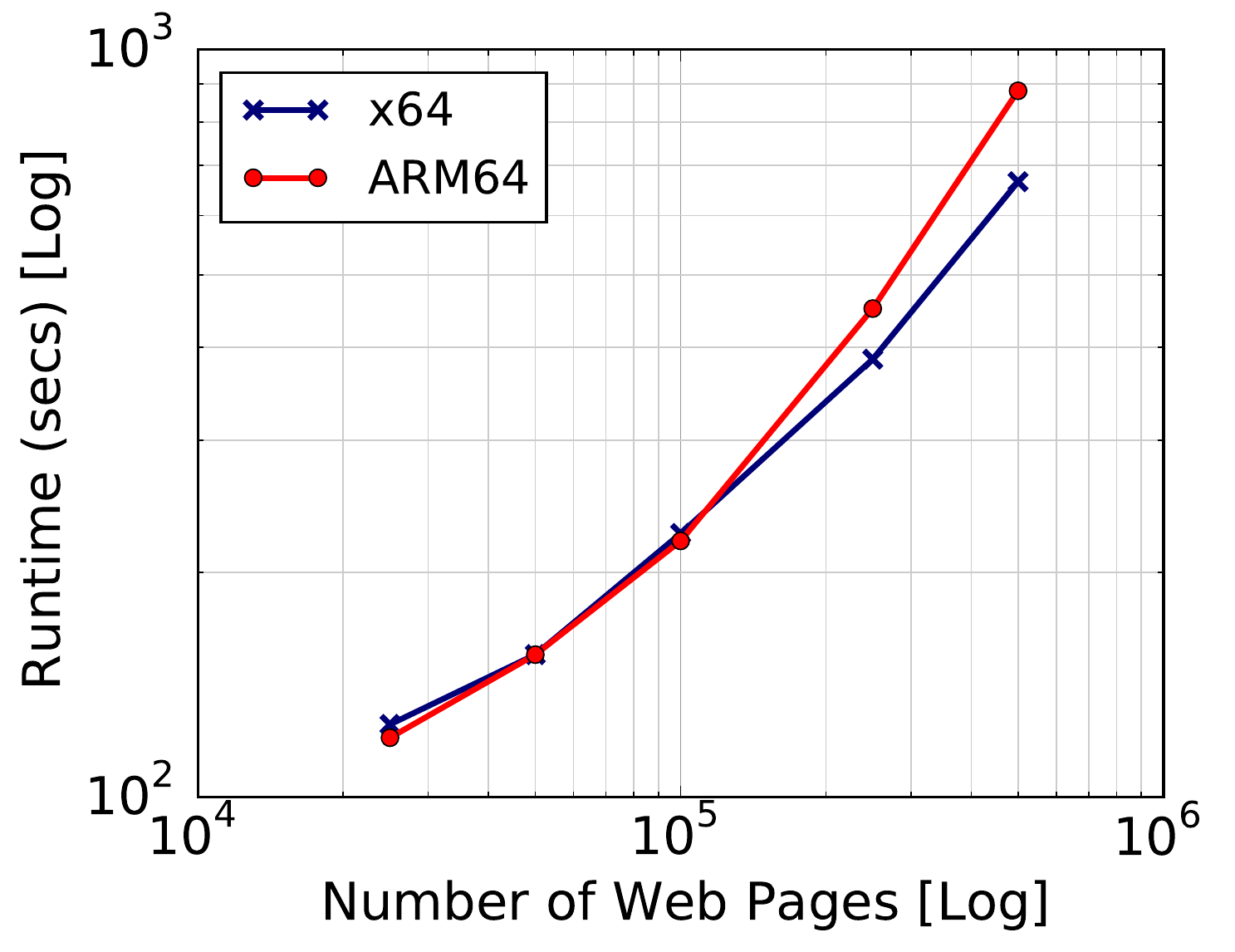}
		\label{fig:time:nutch}
	}
	\caption{Runtime for Web Search Benchmarks}
		\label{fig:time:web}
\end{figure}

\begin{figure}[p]
\centering
        \subfloat[PageRank]{
		\includegraphics[width=0.75\columnwidth]{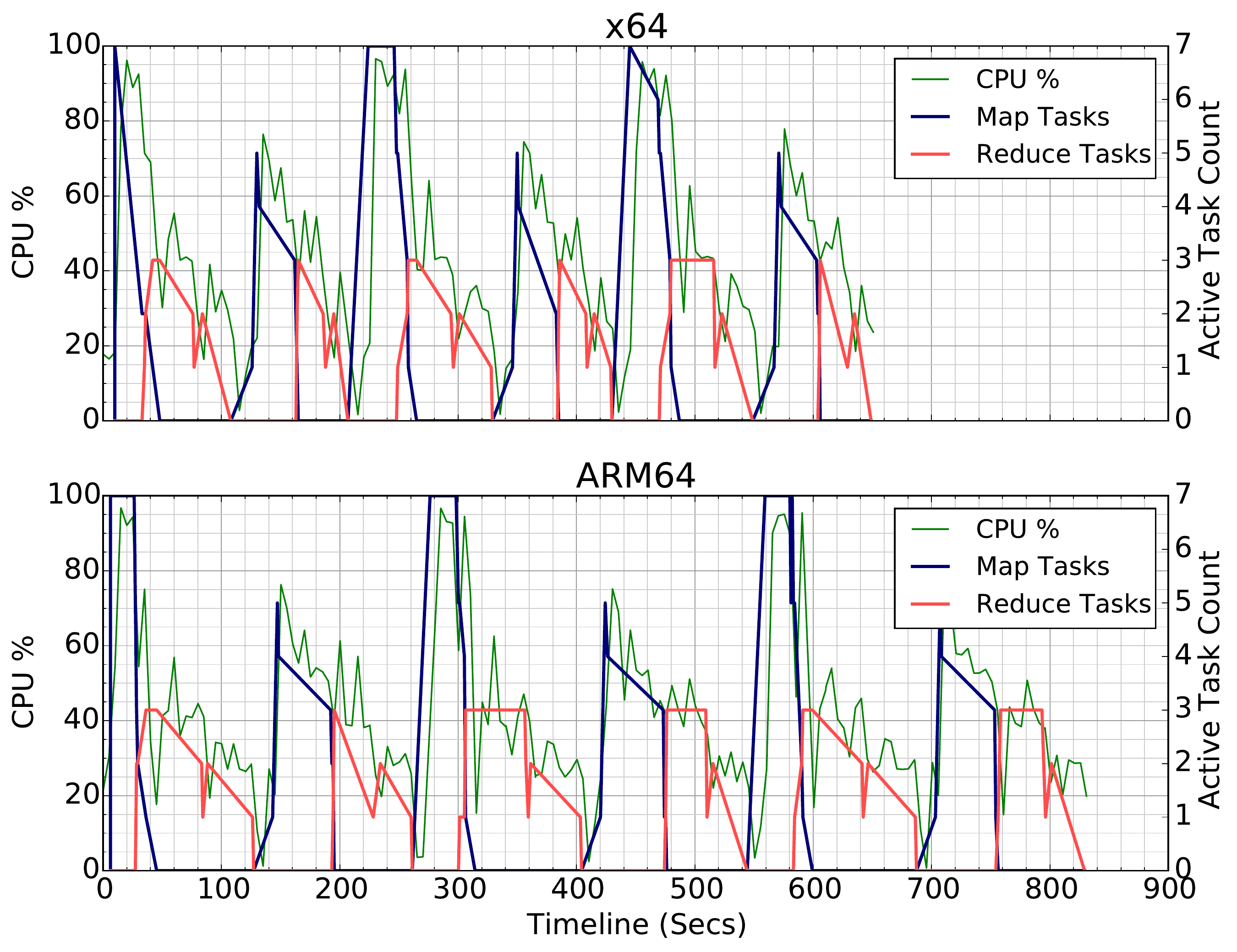}
		\label{fig:cpu:pr}
	}\\
        \subfloat[Nutch Indexing]{
		\includegraphics[width=0.75\columnwidth]{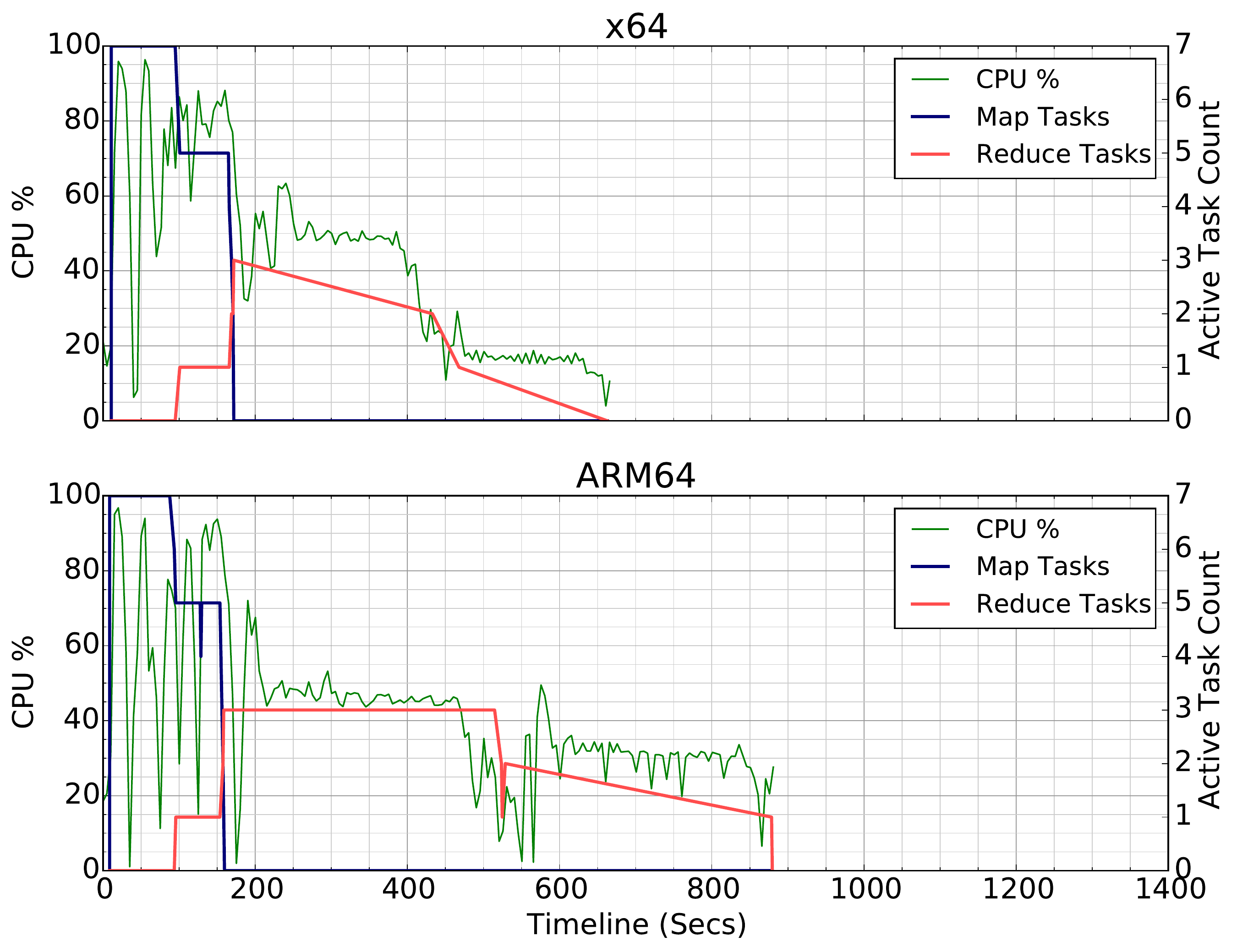}
		\label{fig:cpu:nutch}
	}
	\caption{Timeline plot of active tasks and CPU\% for Web Search Benchmarks for $5 \times 10^5$ pages}
        \label{fig:cpu:web}
\end{figure}

Figs.~\ref{fig:time:web} show the time in seconds taken to perform the PageRank and Nutch Indexing benchmarks, as the number of web pages processed increases on the X axes. Other than the initial small data sizes, the x64 server performs consistently faster than the ARM64 server for both these benchmarks, and this improvement is wider as the data sizes increase. While their runtimes are comparable for $10^5$ pages, this diverges, with the ARM server being slower by 43\% and 32\% for $5 \times 10^6$ and $5 \times 10^5$ pages for the applications. \ysnoted{This translates to an absolute time difference of \note{1921~secs} and \note{370~secs}.}

We drill down into the performance of the Map and Reduce tasks to understand this performance variation. Fig.~\ref{fig:cpu:pr} shows a timeline plot for the duration of the experiment (X axis, seconds) for running \emph{PageRank} on $5 \times 10^5$ web pages. The primary Y axis shows the average CPU usage\% (green line) and the secondary Y axis shows the number of Map (blue) and Reduce (red) tasks active at that time. We see that six MapReduce jobs are executed in waves (spikes in number of tasks), corresponding to pairs of Stage 1 and Stage 2 jobs over 3 iterations in the PEGASUS code for PageRank. \ysnoted{\url{https://goo.gl/ZW0AKM}, Class \texttt{pegasus.PagerankNaive}} We also see that while the relative time taken by Map tasks are comparable for ARM64 and x64, the time taken by the Reduce tasks is tangibly longer for ARM64. This is because the Map tasks perform only integer operations such as string parsing while the Reduce tasks perform floating-point (FP) operations to calculate the rank for each webpage. The ARM Cortex A57 architecture's floating-point performance has previously been shown to be slower than an Intel x64 architecture~\cite{Laurenzano-europar-2014}, and hence this explains its slower performance for PageRank. 
\ysnoted{run linpack to measure FP performance for ARM, x64? Is it possible to pull out just FPU utilization to see if it is bottleneck?}

A similar timeline plot for Nutch, Fig~\ref{fig:cpu:nutch}, shows that the maximum-possible 7 Map tasks run concurrently for the job. 
Here, ARM's mappers are marginally faster than x64's. 
Within the Reduce logic, the two main operations performed are \emph{scoring} and \emph{indexing}, before emitting the index. Scoring is similar to PageRank and is an FP operation, while the indexing and writing are integer and disk operations. Due to the use of just three Reduce task, the progress of Nutch on the ARM64 server is limited by its FP performance for scoring and the indexing's performance on three cores. 
\ysnoted{and also we see that its disk writes in the Reduce task (not shown) is spiky rather than sustained as for the x64}

\subsection{Floating Point Micro-benchmark}
\begin{figure}
	\centering
	\subfloat[ARM64]{
		\includegraphics[width=0.45\textwidth]{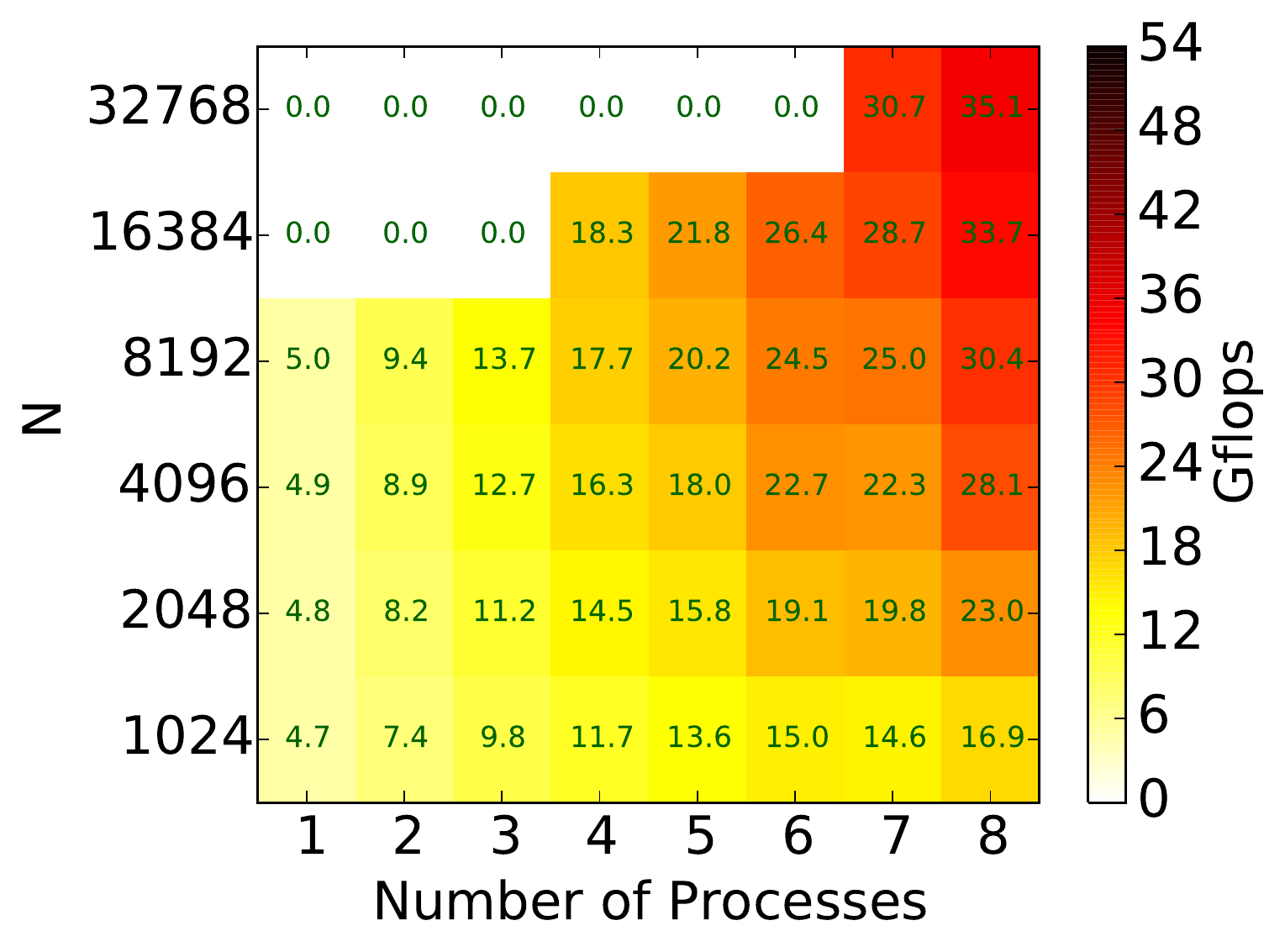}
		\label{fig:heatarm:blis64}
	}
	\subfloat[x64]{
		\includegraphics[width=0.45\textwidth]{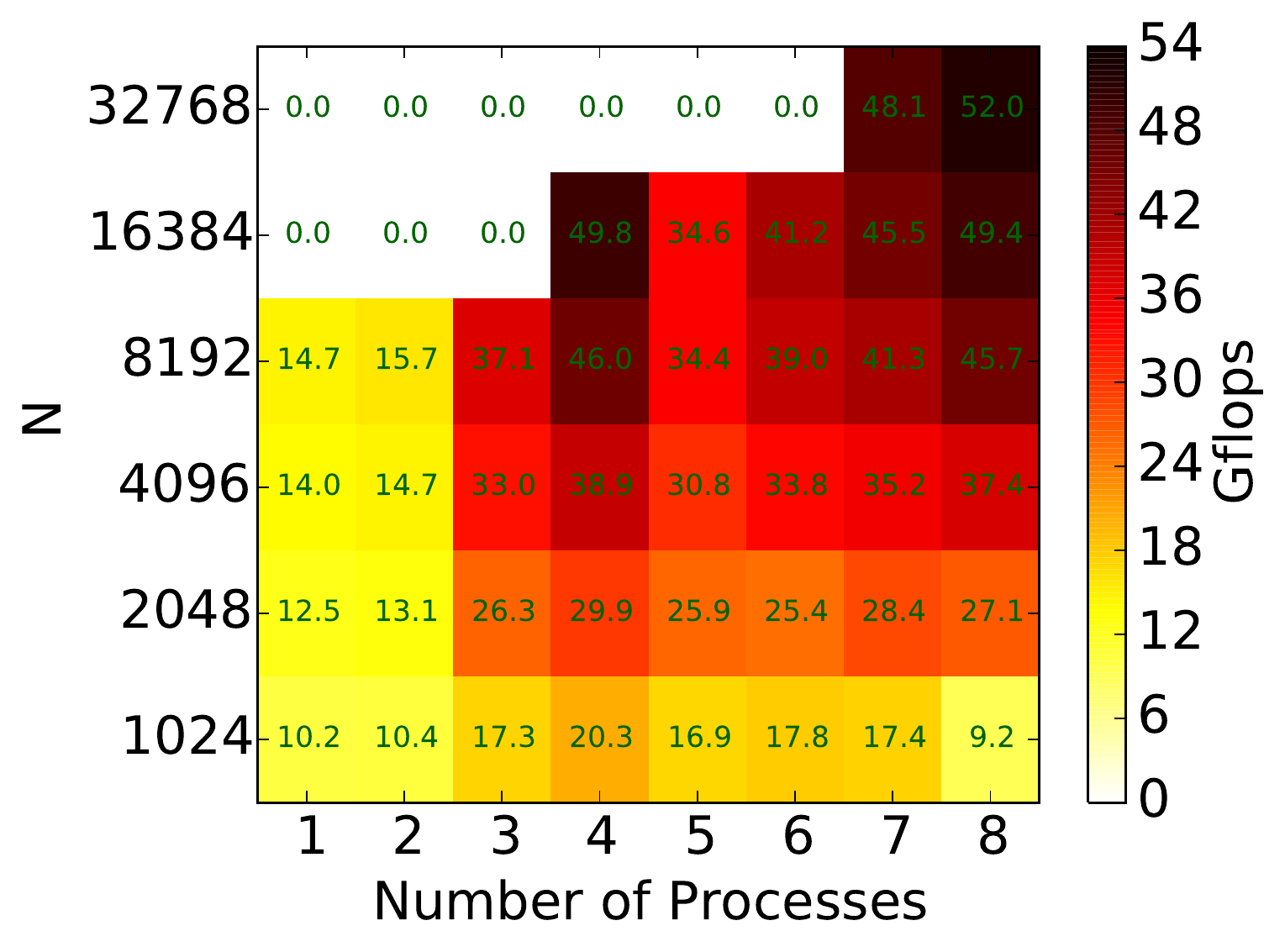}
		\label{fig:heatx64:blis64}
	}
	\caption{HPL performance using BLIS, with Block size = 64}
	\label{fig:heatmap:hpl:blis}
\end{figure}

\begin{figure*}[t]
	\centering
	\subfloat[Scan]{
		\includegraphics[width=0.45\textwidth]{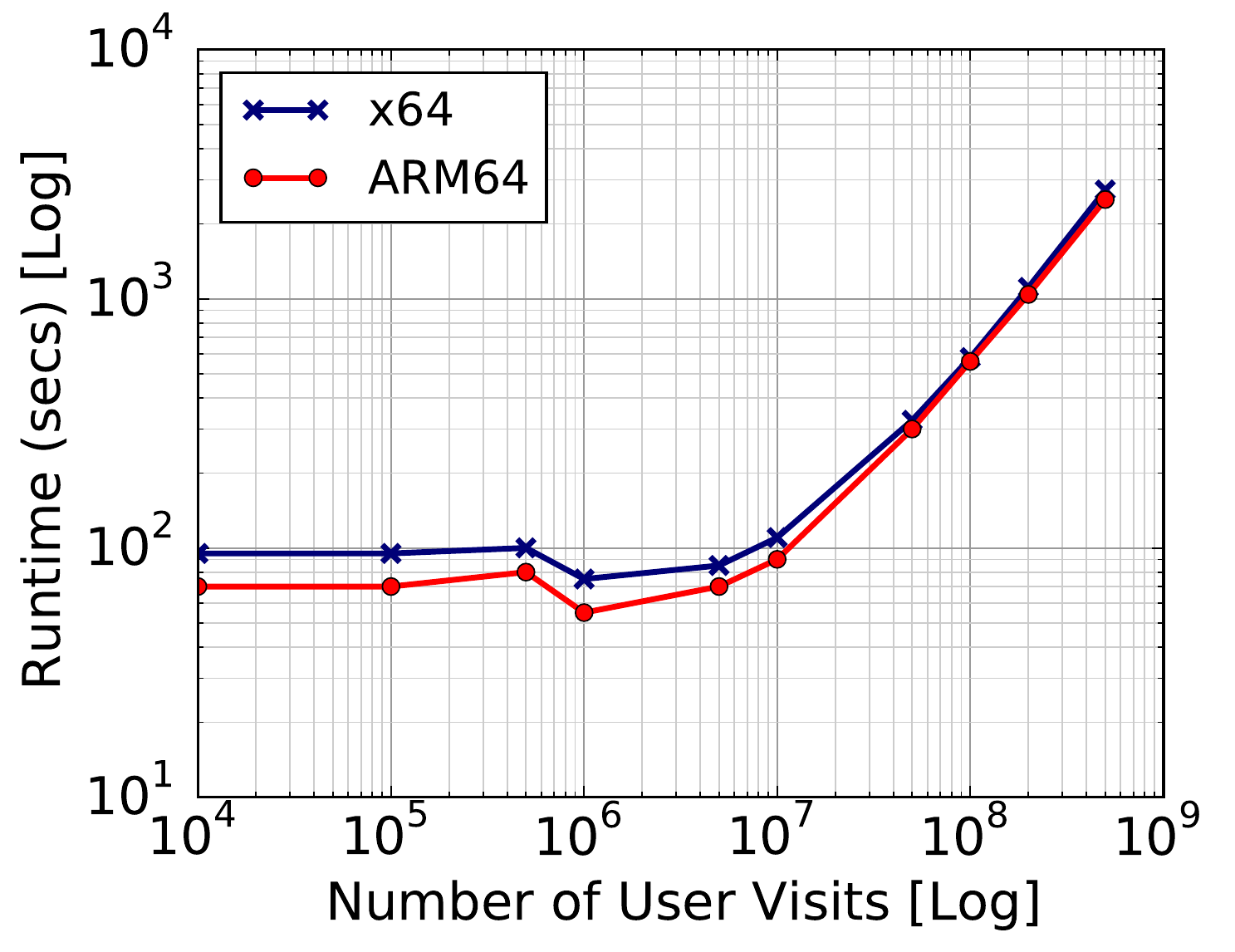}   \label{fig:time:scan}
	}
	\subfloat[Join]{
		\includegraphics[width=0.45\textwidth]{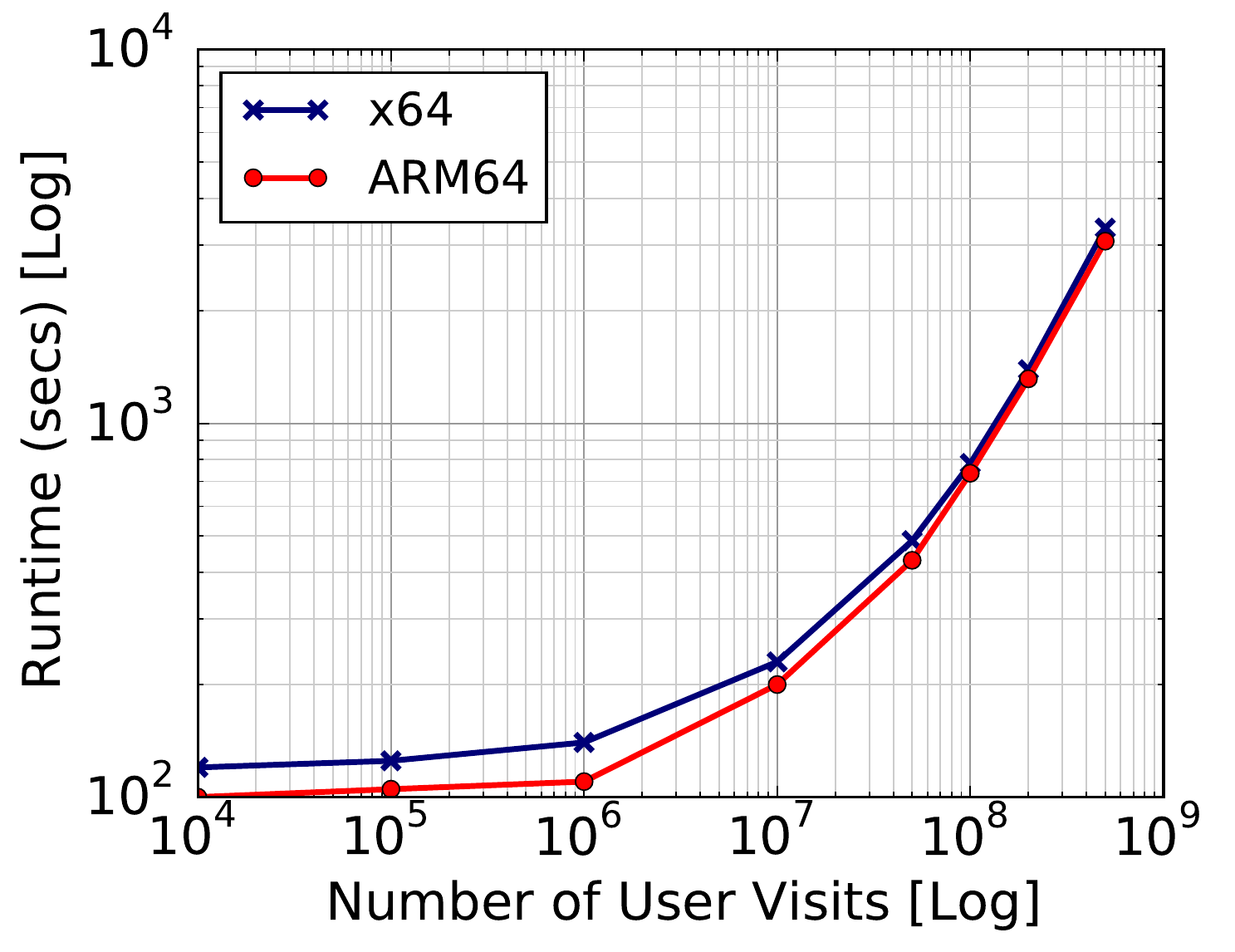}   \label{fig:time:join}	
	}\\
	\subfloat[Aggregation]{
		\includegraphics[width=0.45\textwidth]{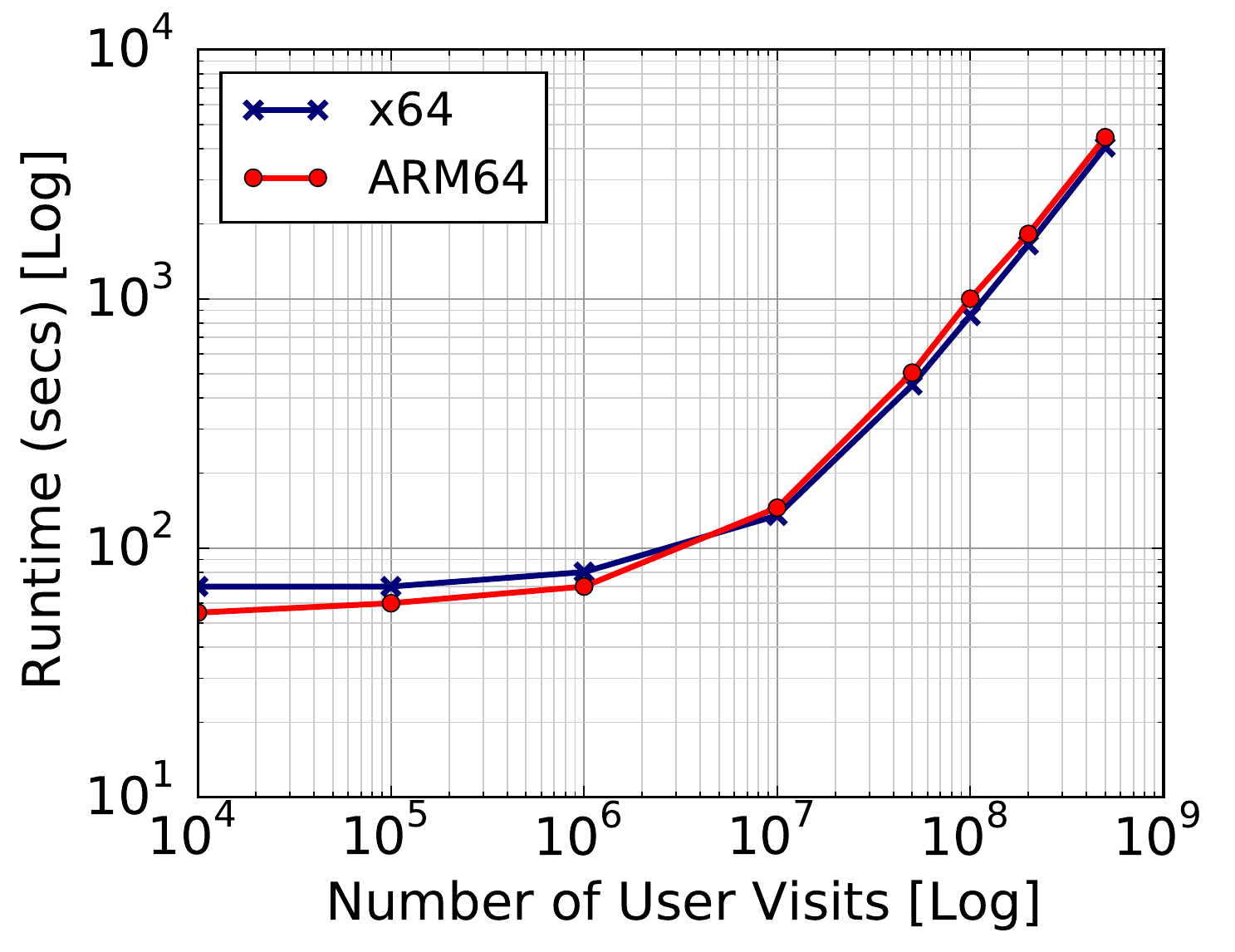}    \label{fig:time:agg}
	}
	\caption{Runtime for Hive Query Benchmarks \ysnoted{X Axis: ``Rows in user visits table''}}
	\label{fig:time:hive}
\end{figure*}

To confirm the hypothesis that slower FP operations on the ARM64 processor are contributing to the slower performance for the web benchmarks, we additionally perform the \emph{High Performance LinPack (HPL)} benchmark~\footnote{http://www.netlib.org/benchmark/hpl/} to evaluate the maximum FP performance on the two processors. 
We use the \emph{BLAS-like Library Instantiation Software (BLIS)} framework implementation which provides architecture-specific optimizations for the ARMv8A and AMD Piledriver micro-architectures used by the two processors, respectively. 
In Fig.~\ref{fig:heatmap:hpl:blis}, we show the heatmap of the performances in GigaFlops in for different matrix sizes (Y Axis)  and number of processes (X Axis), for a fixed block size of 64.
 
We see that x64 achieves 
a peak FP performance $52.3~GFlops$ in comparison to ARM's $35.1~GFlops$. In both cases, this peak value is seen for the most number of processes ($p=8$) and matrix sizes ($n=32,768$) used. 
We note that for a block size of 128 (not shown for brevity), ARM64 managed a peak performance of $42.6~GFlops$ for the same $p$ and $n$, while x64 retains the earlier peak rate but for $p=4$ and $n=16,384$. Even with a block size of 64, we see the x64 
has twin peaks at $p=4$ and $p=8$, while it under-performs for other values of $p$. 
ARM, on the other hand, shows a linear growth in performance as wither $n$ or $p$ increases. 

The x64 Piledriver architecture consists of 4 modules with each having 2 Integer cores and 1 ``Flex FP'' unit shared by those cores~\cite{shilov:2010}. Hence when 4 or 8 processes are run, it allows the system to make best use of all 4 FPU's in the system. In case of the ARM there are 8 FPU's, one per core, and there is a progressive increase in performance as the number of processes reach 8. 

We also notice that the x64 performs $2$--$3\times$ faster on a single process. Some of this difference can be attributed to the clock speeds of $2.0~GHz$ and $2.6~GHz$ for x64 and AMD64, respectively.  
In fact, when we pin the clock speed of the x64 server to $2~GHz$ in a separate HPL benchmark, we see this difference reduce to $1.5$-$2\times$ (plots not shown for brevity). Hence, a single FPU in the x64 is still faster than the ARM64, notwithstanding the clock speeds.

This offers two key takeaways when configuring Big Data platforms like Hadoop for the ARM64 and x64 servers. One, that applications that are FP intensive are in general likely to be slower on ARM64 than x64. Two, that ARM64 can offer linear scaling of FP performance from $1-8$ cores, while the x64 AMD Piledriver has sweet spots when using $4$ or $8$ cores.

\subsection{Hive Query Benchmarks}

\ysnoted{\begin{figure}[t]
	\centering
		\includegraphics[page=2,width=0.75\columnwidth]{lineplot-scan-min9.pdf}      
	\caption{Timeline plot of disk reads at $5~sec$ intervals, when running the \emph{Scan} query on user visits table with $5 \times 10^8$ rows}
\label{fig:reads:scan}
\end{figure}}

The \emph{scan} and \emph{join} queries on Hive perform faster on the ARM64 server than the x64 for tables with fewer rows, but the advantage narrows (but is still present) for larger sized tables. 
This is shown in Figs.~\ref{fig:time:scan} and \ref{fig:time:join}. E.g., for the largest table size, scan takes $2,502~secs$ on ARM64 and $2,733~secs$ on x64, while join takes $3,073~secs$ and $3,329~secs$ on the two servers, respectively. On the other hand, for the \emph{aggregation} query, ARM64 starts off faster but is overtaken by the x64 server for larger table sizes, as we see in Fig.~\ref{fig:time:agg}. Here, the corresponding times for ARM64 and x64 times for the largest data size are $4,449~secs$ and $4,064~secs$. Two factors contribute to this.

First, all Hive queries are compiled into an execution plan of MapReduce job(s), which takes some initialization time~\footnote{Apache Hive EXPLAIN Syntax, \url{https://cwiki.apache.org/confluence/display/Hive/LanguageManual+Explain}}. 
This time is consistently faster on the ARM64 server than x64 by about $10~secs$. E.g., ARM64 takes $\sim44~secs$ to plan a join query while x64 takes $\sim54~secs$.
\ysnoted{For e.g., these pairs of times in seconds for $\langle$ARM64, x64$\rangle$ servers for the scan, join and aggregate queries are $\langle 33\sim38, 44\sim50\rangle$, $\langle 42\sim46, 53\sim55\rangle$, and $\langle 32\sim35, 40\sim44 \rangle$, respectively.} These absolute values are significant for smaller data sizes.

\ysnoted{Read and write bytes for ARM and x64 are different. ARM does 10-30\% more bytes read and write for aggregation, 20\% more writes and 7\% less reads for join. Block sizes are different, 8k for ARM, 4k for x64. Could that explain? Check with Hortonworks?}

Second, the performance of the aggregate query is slower on ARM64 for larger dataset sizes due to its use of the \texttt{sum()} FP function in both Map and Reduce logic, 
though the bulk of rows for this operation is performed is in the Reduce task. 
Consequently, the poor FP performance of the ARM64 server is to blame, 
but the execution planning time works in ARM's favor to mitigate this impact.

\subsection{Machine Learning Benchmarks}

\begin{figure}[t!]
	\centering
	\subfloat[Na\"{i}ve Bayes Classifier Training]{
		\includegraphics[width=0.45\columnwidth]{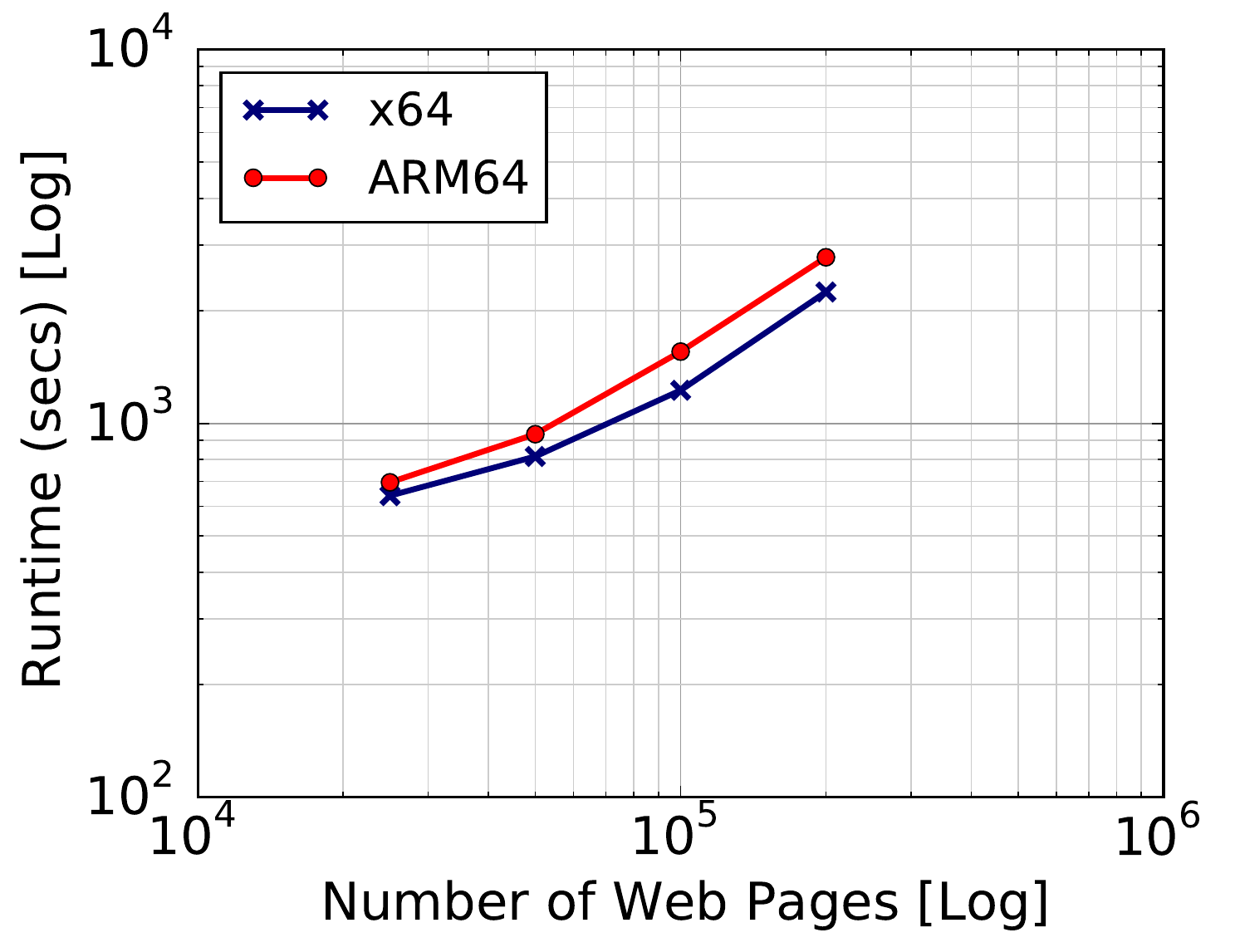}
		\label{fig:time:bayes}
	}
	\subfloat[K-Means Clustering \ysnoted{X Axis: ``Number of Web Pages''}]{
		\includegraphics[width=0.45\columnwidth]{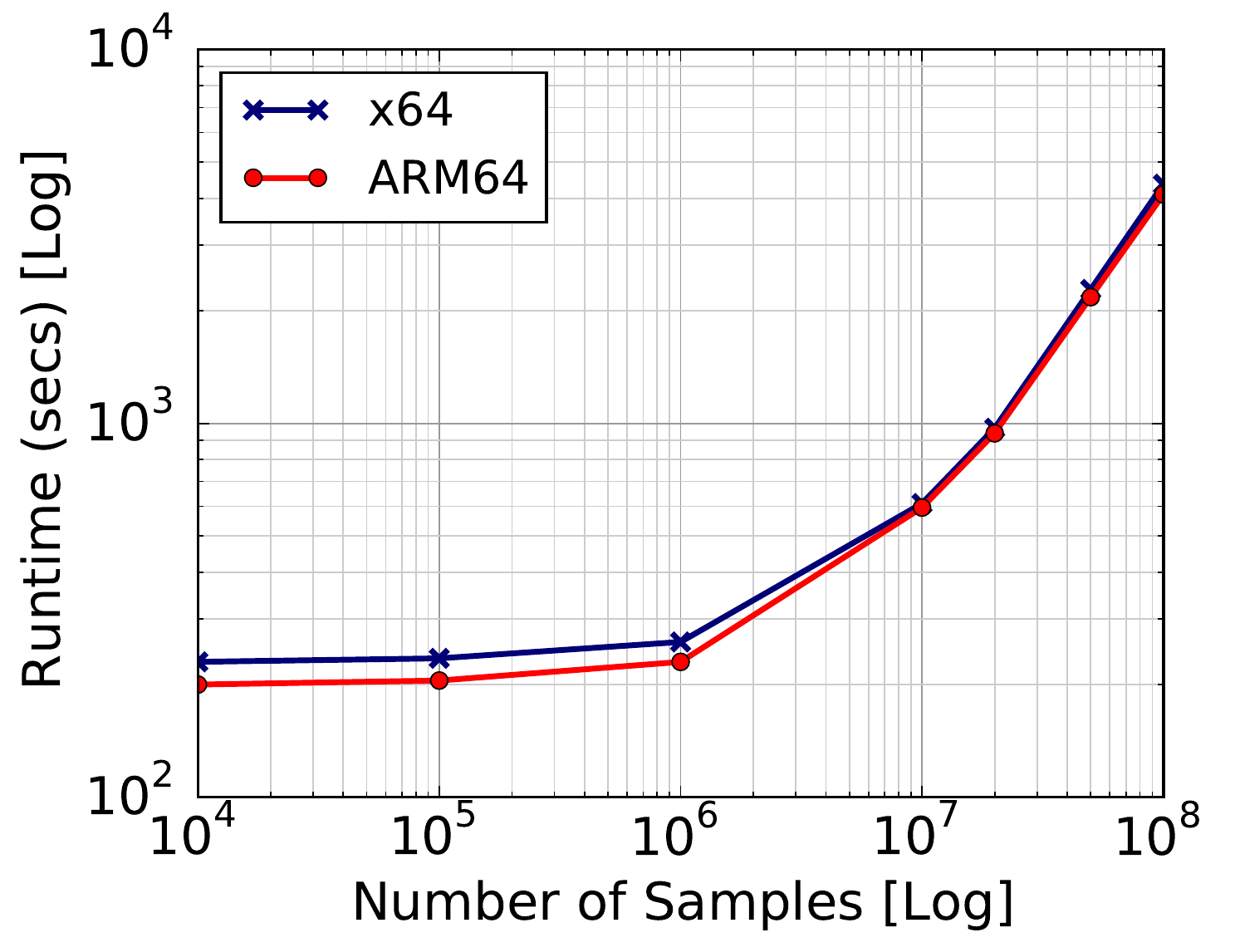}
		\label{fig:time:kmeans}
	}
	\caption{Runtime for Machine Learning Benchmarks}
        \label{fig:time:ml}
\end{figure}

\subsubsection{Na\"{i}ve Bayes Classifier Training}
The runtime for Bayes classifier training is shown in Fig.~\ref{fig:time:bayes} for different training data sizes. We observe that as the number of input web pages grows from $25\times10^3$ to $2\times10^5$, the x64 server outperforms the ARM in all cases, with the x64 taking $\sim19\%$ lesser time for the largest data sizes. Looking at a timeline plot of the active Map and Reduce tasks, and the CPU\% in Fig.\ref{fig:cpu:bayes}, we see that second MapReduce job dominates the total time (the first being a Map-only job). In particular, the Reduce task for ARM is much slower than the x64 server. In both cases 3 Reduce tasks run concurrently and consume about $20\%$ of CPU. Here, the logic that is the bottleneck is the \texttt{CollocReducer} class, which pre-dominantly performs Vector integer operations, that appear slower on ARM64.

\subsubsection{K-Means Clustering}
\begin{figure}[p]
	\centering
	\subfloat[Na\"{i}ve Bayes Training on {$10^{5}$} pages]{
		\includegraphics[width=0.75\columnwidth]{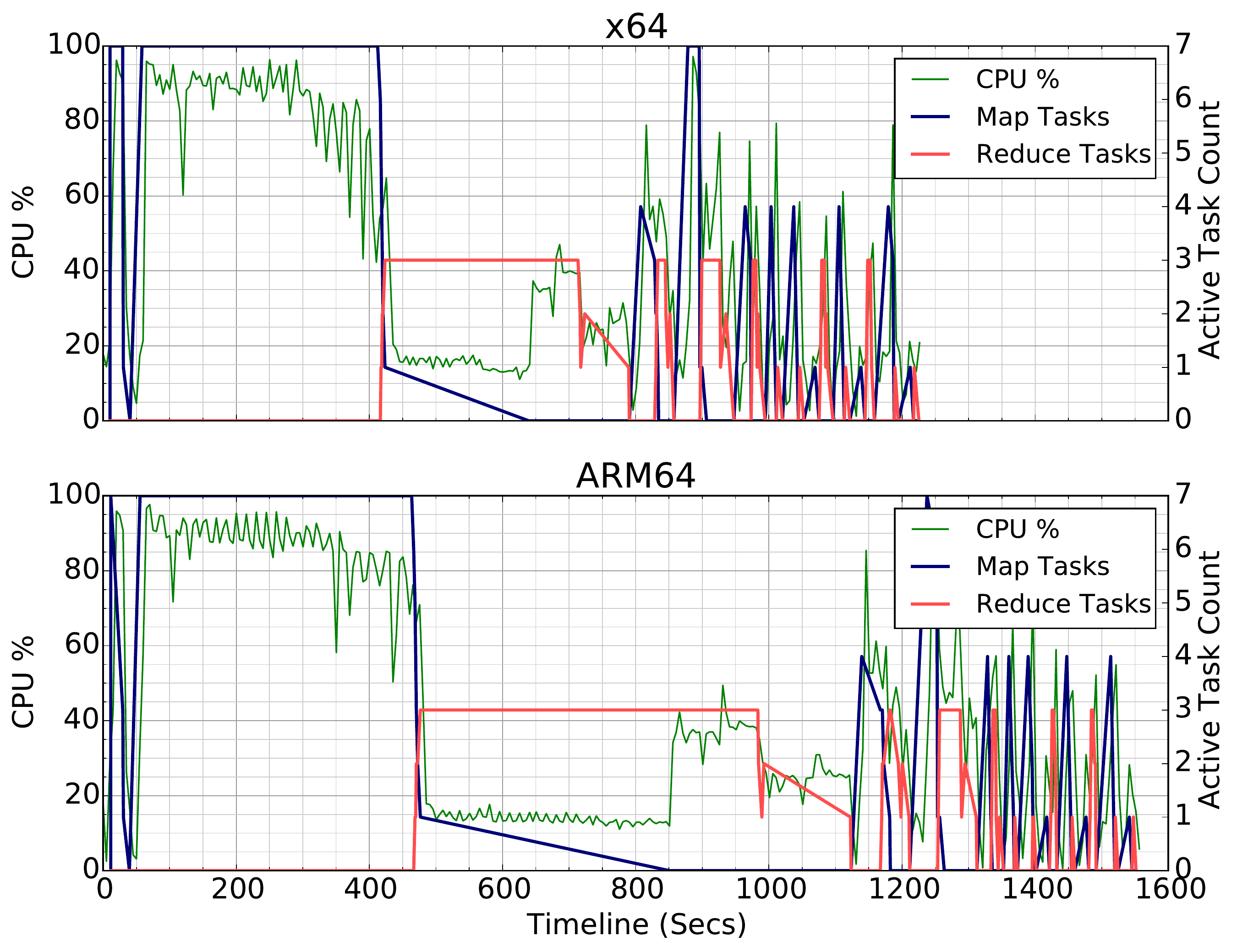}
		\label{fig:cpu:bayes}
	}\\
	\subfloat[K-Means Clustering on {$5\times 10^{7}$} pages]{
		\includegraphics[width=0.75\columnwidth]{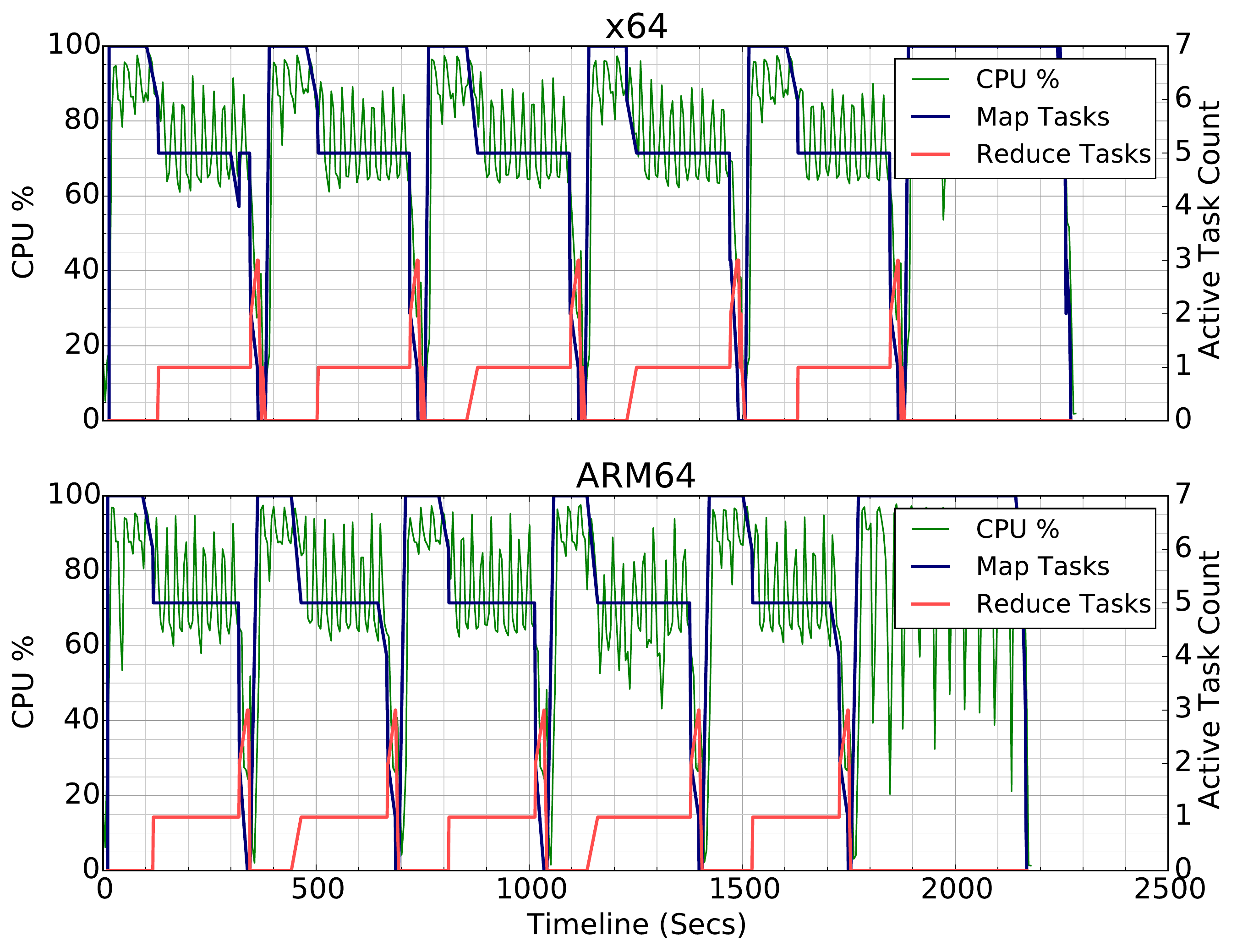}
		\label{fig:cpu:kmeans}
	}
	\caption{Timeline plot of active tasks and CPU\% for ML} 
        \label{fig:cpu:ml}
\end{figure}

The ARM64 server performs K-Means Clustering faster than the x64 server, by around $20-26~secs$ for smaller workloads of sizes $10^4-10^6$ (Fig.~\ref{fig:time:kmeans}). However, this advantage does not hold for larger data sizes and both servers' performances are on par with each other. Interestingly, K-Means makes use of heavy FP operations in both Map and Reduce tasks, e.g., to identify distance of points from cluster centroids and to test if the algorithm has converged.
The timeline plots for the clustering (Fig~\ref{fig:cpu:kmeans}) show that at almost every timepoint, the maximum of 7 possible tasks are active, Map or Reduce. In fact, Reduce tasks completely overlap with the Map tasks, and when they do, the number of active Map tasks drops. Consequently, the full CPU parallelism is exploited by both servers and ARM performs almost as fast as x64.
\ysnoted{since we use default reducer count config, only 1 reducer is active}

\subsection{Reducer Parallelism Tuning}

We see that for several applications that are floating-point heavy in the Reduce tasks, the number of Reduce tasks are lesser than the number of cores since the higher memory allocation for Reduce containers limits the number of concurrent Reduce tasks to 3. Consequently, only 3 of the 8 FPUs available in the ARM cores are utilized by Reduce tasks, and only 3 of the 4 shared FPUs are used on the Opteron.

To confirm this hypothesis, we reduce the memory allocated to the Reduce containers to $1920~MB$ to allow up to 7 concurrent Reducers to run. 
We rerun  PageRank, Nutch, and Aggregation applications that showed poorer performance on ARM using this configuration for one large dataset. \ysnoted{, and Bayes}

\begin{figure}[p]
	\centering
        \subfloat[PageRank]{
		\includegraphics[width=0.75\columnwidth]{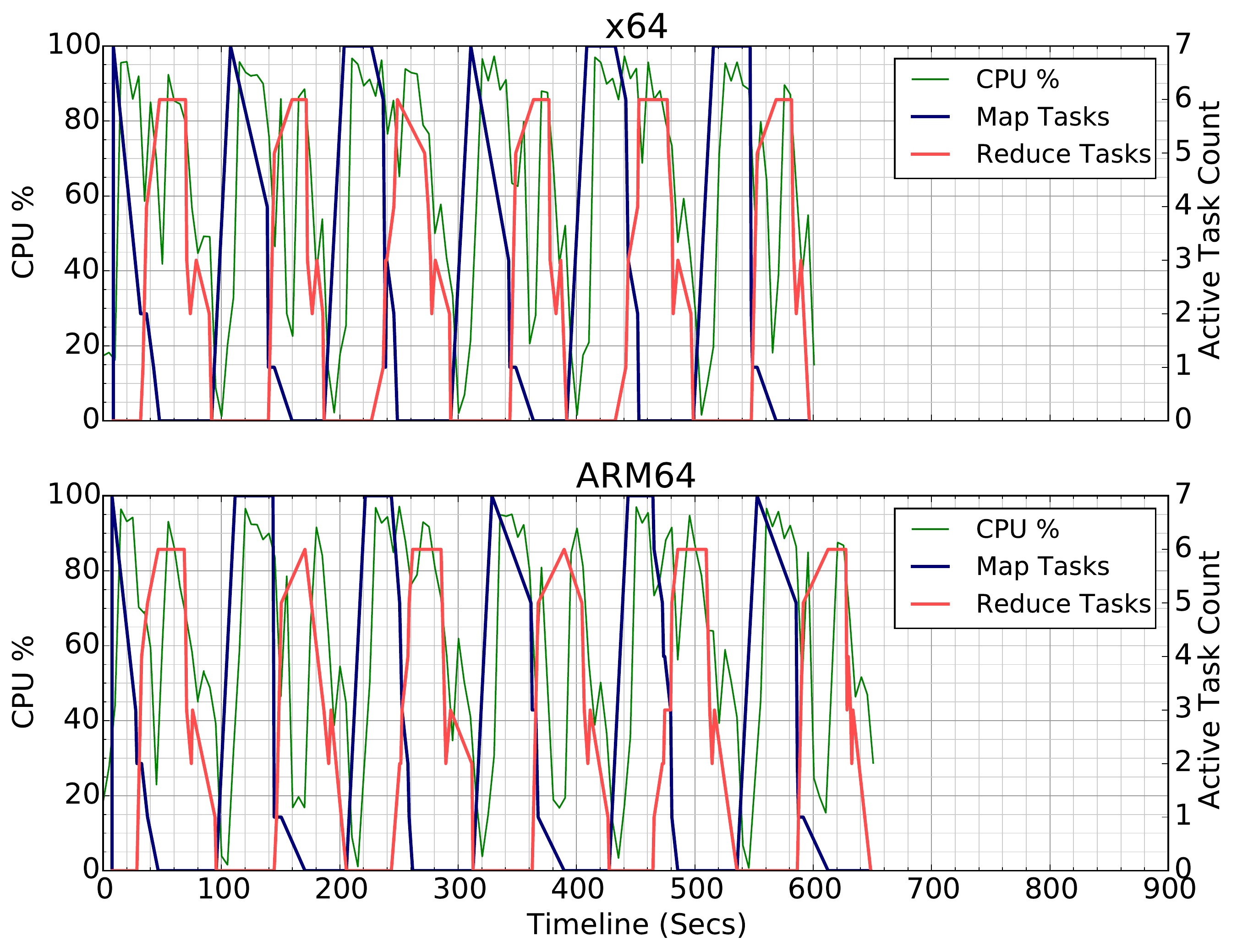}
		\label{fig:cpu:pr:8R}
	}\\
        \subfloat[Nutch Indexing]{
		\includegraphics[width=0.75\columnwidth]{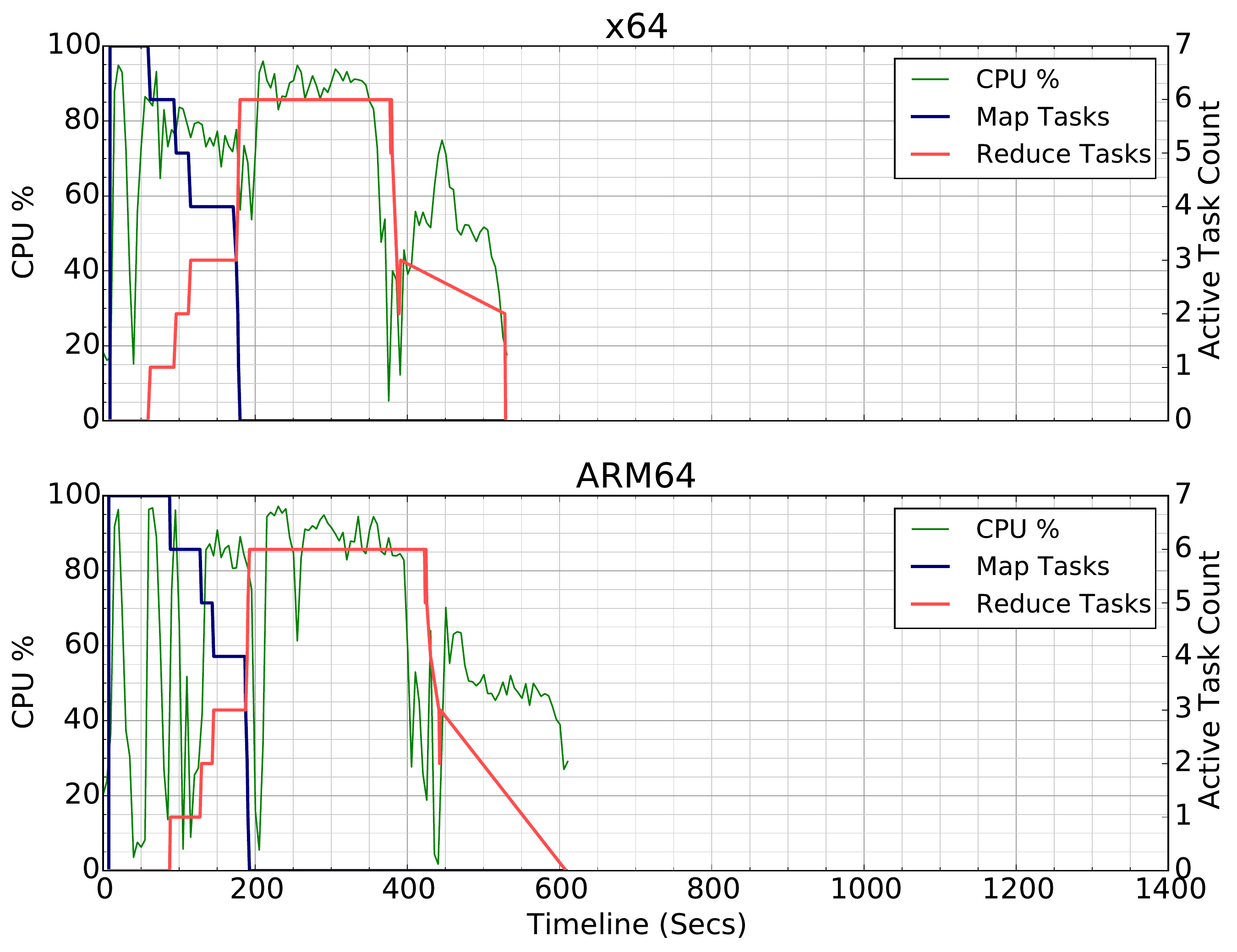}
		\label{fig:cpu:nutch:8R}
	}
	\caption{Timeline plot of active tasks and CPU\% for Web Search Benchmarks for $5 \times 10^5$ pages, with \emph{higher Reducer parallelism}}
\end{figure}
For \emph{PageRank} with an input data having $5 \times 10^5$ pages, we see the runtime for the ARM server drops from $830.9~secs$ earlier to $650.7~secs$ with the low-memory, but more number of Reduce tasks (Fig.~\ref{fig:cpu:pr:8R}). 
The x64 server also saw a modest drop from $650.8~secs$  to $600.7~secs$.
Earlier, the 3 Reduce tasks active for PageRank could use $\frac{3}{8}^{th}$ of the FPUs in ARM and $\frac{3}{4}^{th}$ for x64, while in the new setup, the 7 Reduce tasks use $\frac{7}{8}^{th}$ of the ARM FPUs and all 4 of the x64's FPUs. So the relative improvement for ARM is much more than x64, and the performance difference narrows down to under $8\%$. Some of the improvement is due to additional CPU cores as well, as seen by the higher CPU\%.

When we increase the number of Reducers for the Nutch benchmark, we see from Fig.~\ref{fig:cpu:nutch:8R} that the runtime drops by a third, with up to $6$ active Reduce tasks, relative to the $3$ Reducer setup in Fig.~\ref{fig:cpu:nutch}, and the ARM64 server almost matches x64 on the runtime.
Similar improvements were observed for Aggregation, but its discussion omitted due to space constraints.

\ysnoted{\begin{figure}
  \centering
\subfloat[Default Reducer Parallelism]{
		\includegraphics[width=0.75\columnwidth]{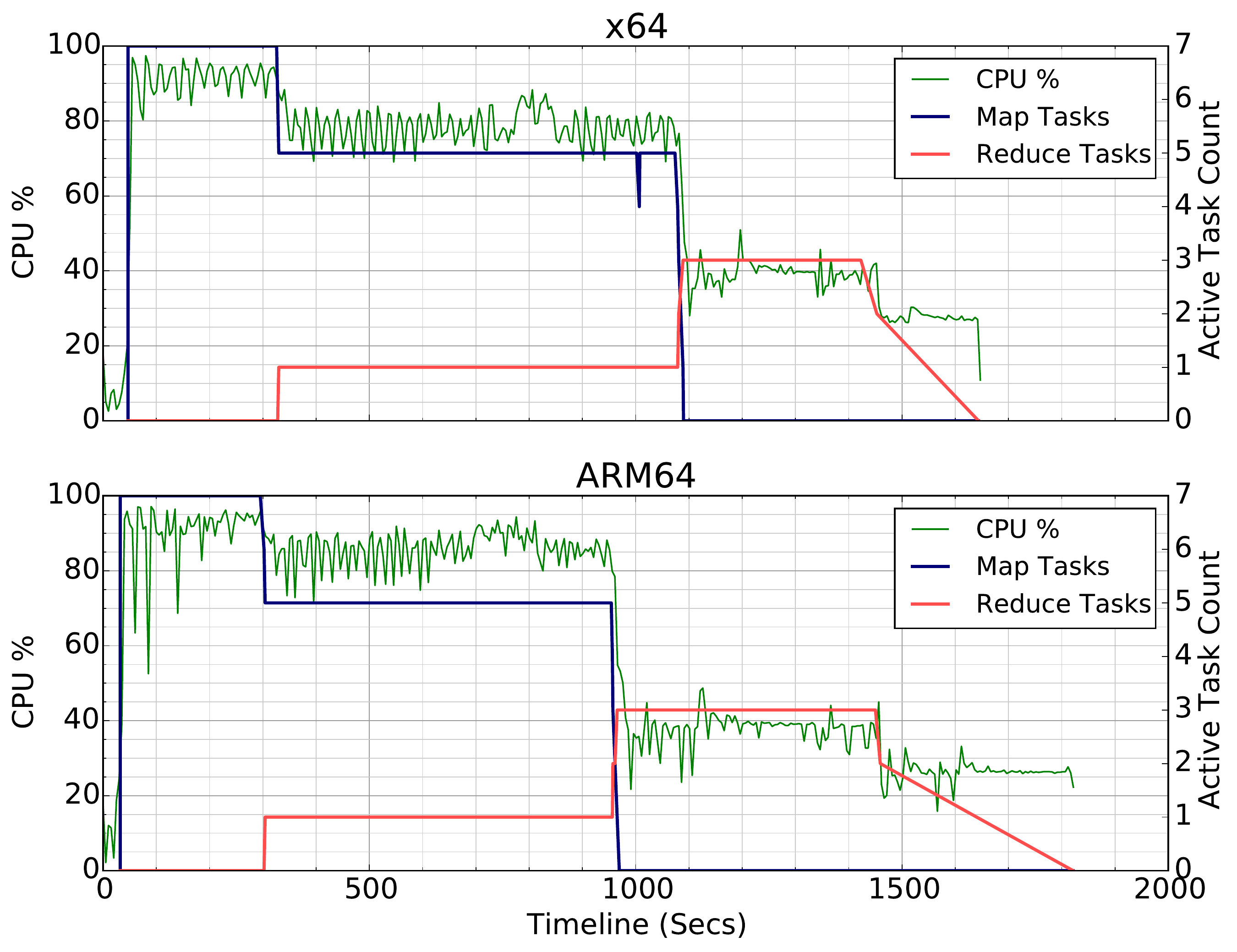}
		\label{fig:cpu:agg:4R}
	}\\
\subfloat[Higher Reducer Parallelism]{
		\includegraphics[width=0.75\columnwidth]{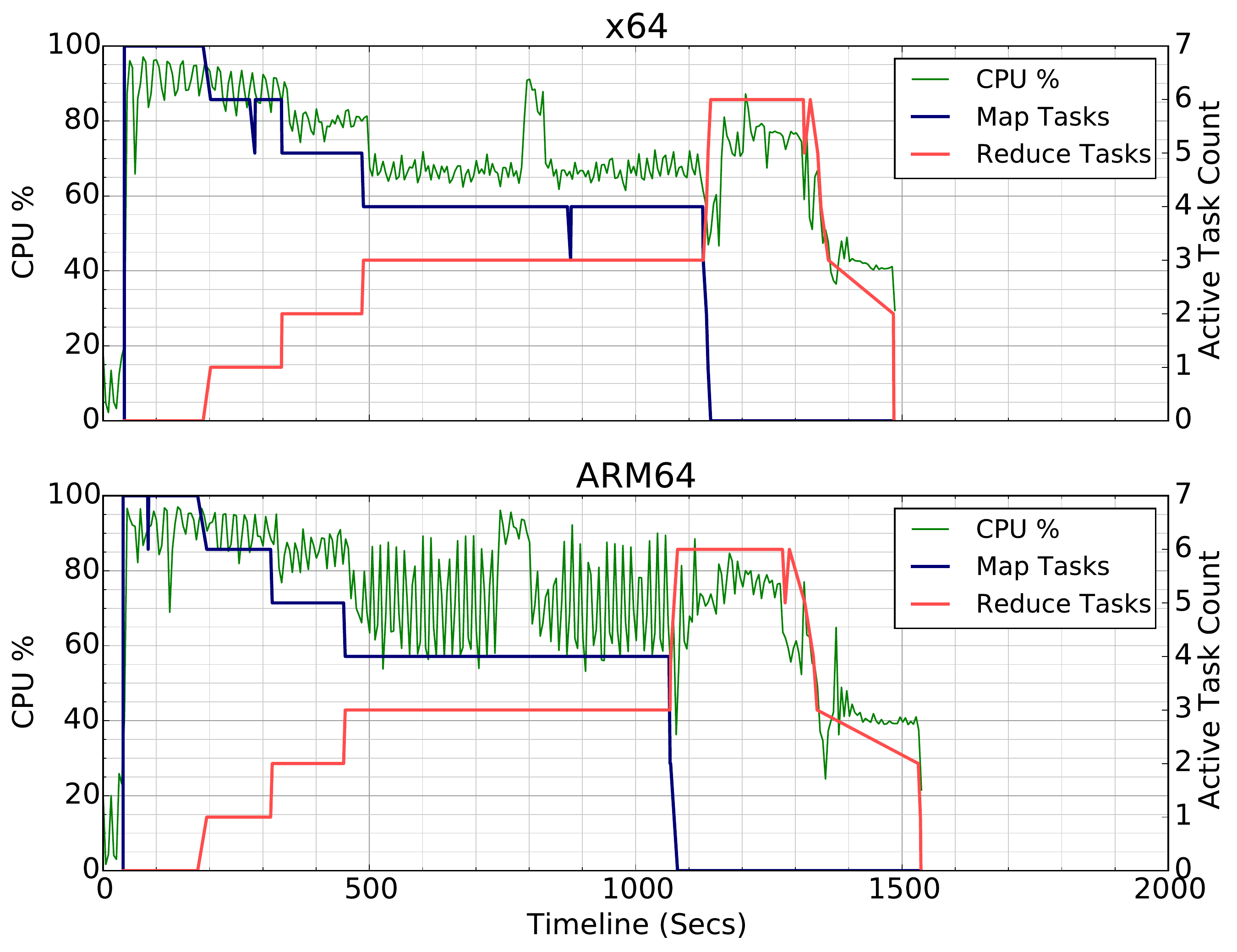}
		\label{fig:cpu:agg:8R}
	}
  \caption{Timeline plot of active tasks and CPU\% for Aggregation query for $2 \times 10^8$ input rows}
  \label{fig:cpu:agg}
\end{figure}

When we run the \emph{aggregation} query with more numbers of reducers for an input data size of $2\times10^8$, we observed a tangible reduction in the runtime for ARM64 relative to having fewer reducers earlier, dropping from $1,821~secs$ to $1,536~secs$ (Figs.~\ref{fig:cpu:agg}). This reduced time is almost comparable to the time taken by x64, which drops by $\sim9\%$ to $1,486~secs$. The key reason for this is the use of up to 7 Reduce tasks in the ARM64 with corresponding floating-point parallelism to perform the \texttt{sum} aggregation. We also notice that having the additional Reducers allows the job to start the Reduce tasks much before the Map tasks complete, giving a longer overlap of the Map and Reduce phases. Of course, this also has the downside of slowing down the Map phase in ARM64, which now takes almost as long as the x64 to complete. 
}
\ysnoted{Reduce does the sum(FP), so is slower on ARM. Map part is faster on ARM. Why? Lot of 100MB reads and 50MB writes in the Map phase. Lot of 100MB writes in Reduce.}
\ysnoted{more reducers also means early start of reducers overlapping with mappers}

\ysnoted{For the \emph{Bayes benchmark}, the improvement gained by increasing the parallelism of reduce tasks was more modest in the ARM64 while in case of x64 there was a small increase in the taken taken to complete. }
\ysnoted{Bayes suffers from slow reduce on ARM. But increasing tasks does not help. Not clear why reduce is much slower, Map is slightly slow. Lot of 100MB writes in Map phase, 20MB writes in reduce phase}
\ysnoted{CollocReducer: TODO...Explain!}

\section{Energy Efficiency Analysis}
\label{sec:energy}
Neither server exposes on-board energy counters or power profiling using DCMI. So we sample the instantaneous energy consumed by the servers (in Watts) when running the above benchmarks, every $20~secs$ using a load measurement device
~\footnote{Joule Jotter: Collecting power utilization datasets from Households and Buildings, \url{http://homepage.tudelft.nl/w5p50/jj/}}, which is placed between the power socket and each server.

Also, the power supply units (PSU) to the two servers are not similar. The Overdrive 3000 ARM64 server is a 1U blade with an independent PSU rated at $200~Watts$ ($W$). The Opteron 3380 x64 server is a thin blade in a single 12-node 3U Enterprise chassis, with a pair of redundant Platinum efficiency PSUs rated at $1620W$ that is shared by all 12 blades. To ensure fair comparison with the ARM server, we power off $11$ of the $12$ blades, and also power off the redundant PSU. 

We measure the base power consumed by the servers, when they are freshly booted and idle, over a $2~hour$ period.
This average idle power (base load) is $45.30W$ for the ARM64 server and $134.14W$ for the x64 server.

\begin{figure}[t!]
	\centering
	\includegraphics[width=0.50\columnwidth]{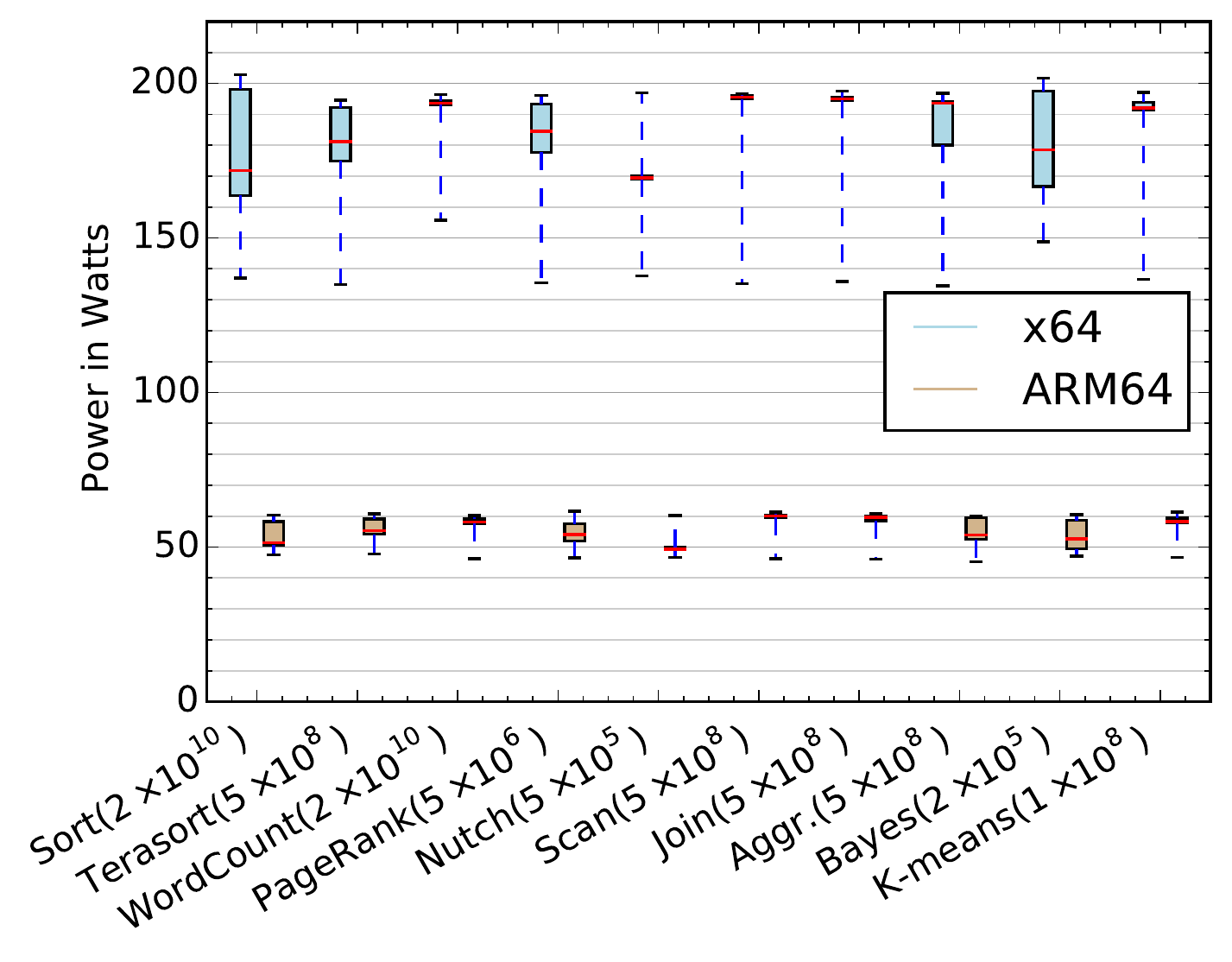}
	\caption{Average power consumed ($W$) for all benchmarks for their largest workload size (labeled in X Axis)}
	\label{fig:pow:all}
\end{figure}

Fig.~\ref{fig:pow:all} shows the box plot for the average power consumed over the $20~sec$ samples measured when each benchmark was running for the largest dataset size. We omit DFSIOe since we only use it to measure the HDFS throughput. We can see that the median power consumption for all benchmarks falls within a narrow band for each server. This ranges from $50$--$60~W$ for ARM64 and $170$--$193~W$ for x64. This shows that the base load dominates and the incremental load for the benchmarks themselves is only $\sim 33\%$ more than the base in ARM and $\sim 44\%$ more than the base in x64.
We also notice that high CPU usage applications such as Word Count and K-means have a higher incremental power consumption compared to lower CPU usage applications such as Nutch indexing and Bayes which indicates a correlation between CPU usage and power consumption

\begin{figure}[t!]
	\centering
	\includegraphics[width=0.85\columnwidth]{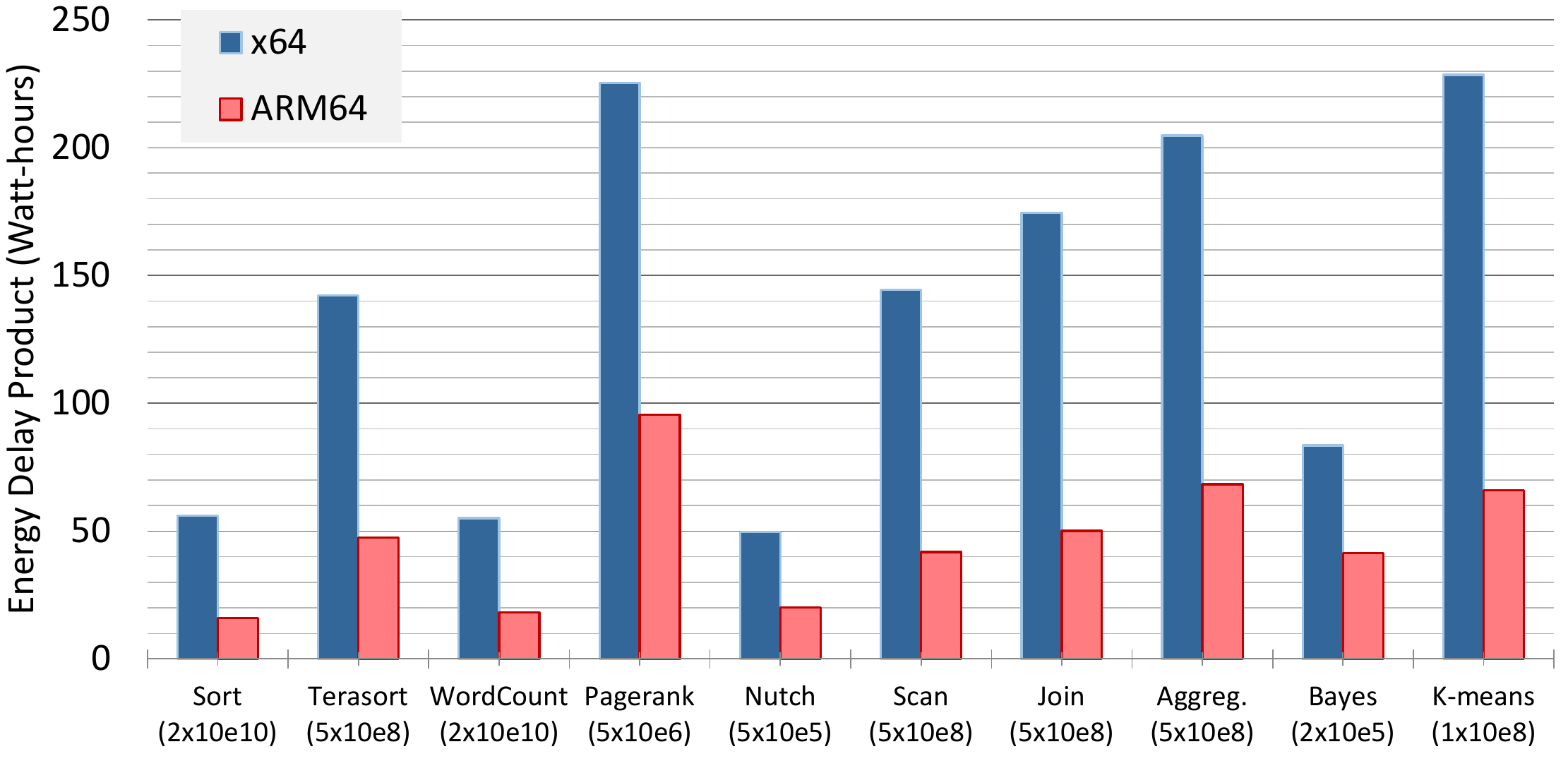}
	\caption{Energy Delay Product (EDP, in $Wh$) for all benchmarks for their largest workload size (labeled in X Axis)}
	\label{fig:edp:all}
\end{figure}
The \emph{Energy Delay Product (EDP)} is a common metric to evaluate the effective energy consumption for applications and benchmarks~\cite{edp}, and offers a measure of the operational cost for purchasing power from the utility~\cite{rodero2010energy}. We calculate this for our experiments by multiplying the average power load sampled with the runtime duration of the benchmark. Since the sampling interval is a fixed $20~secs$, this is a reasonable estimate of EDP. Fig.~\ref{fig:edp:all} shows this for the largest workload sizes for each benchmark, in $Watt-hours (Wh)$. Here, the difference between the ARM64 and the x64 servers is stark. ARM has a $50-71\%$ lower EDP than x64, thus verifying that the ARM64 has a much more favorable mix of both performance and energy efficiency.

\begin{figure}[t!]
	\centering
        \subfloat[Nutch Indexing]{
		\includegraphics[page=1,width=0.45\columnwidth]{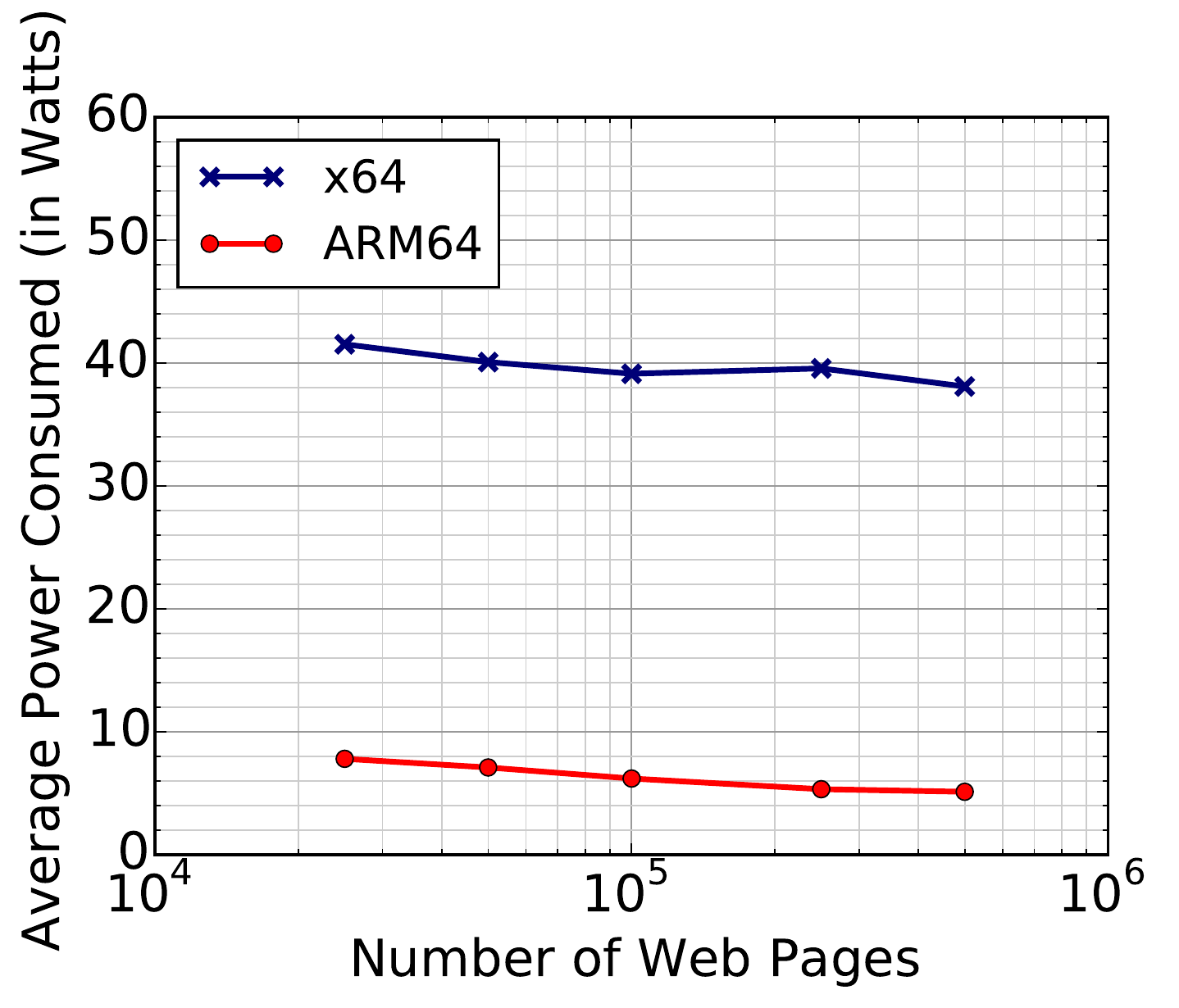}
		\label{fig:incpow:nutch}
	}
	\subfloat[K-means]{
		\includegraphics[page=1,width=0.45\columnwidth]{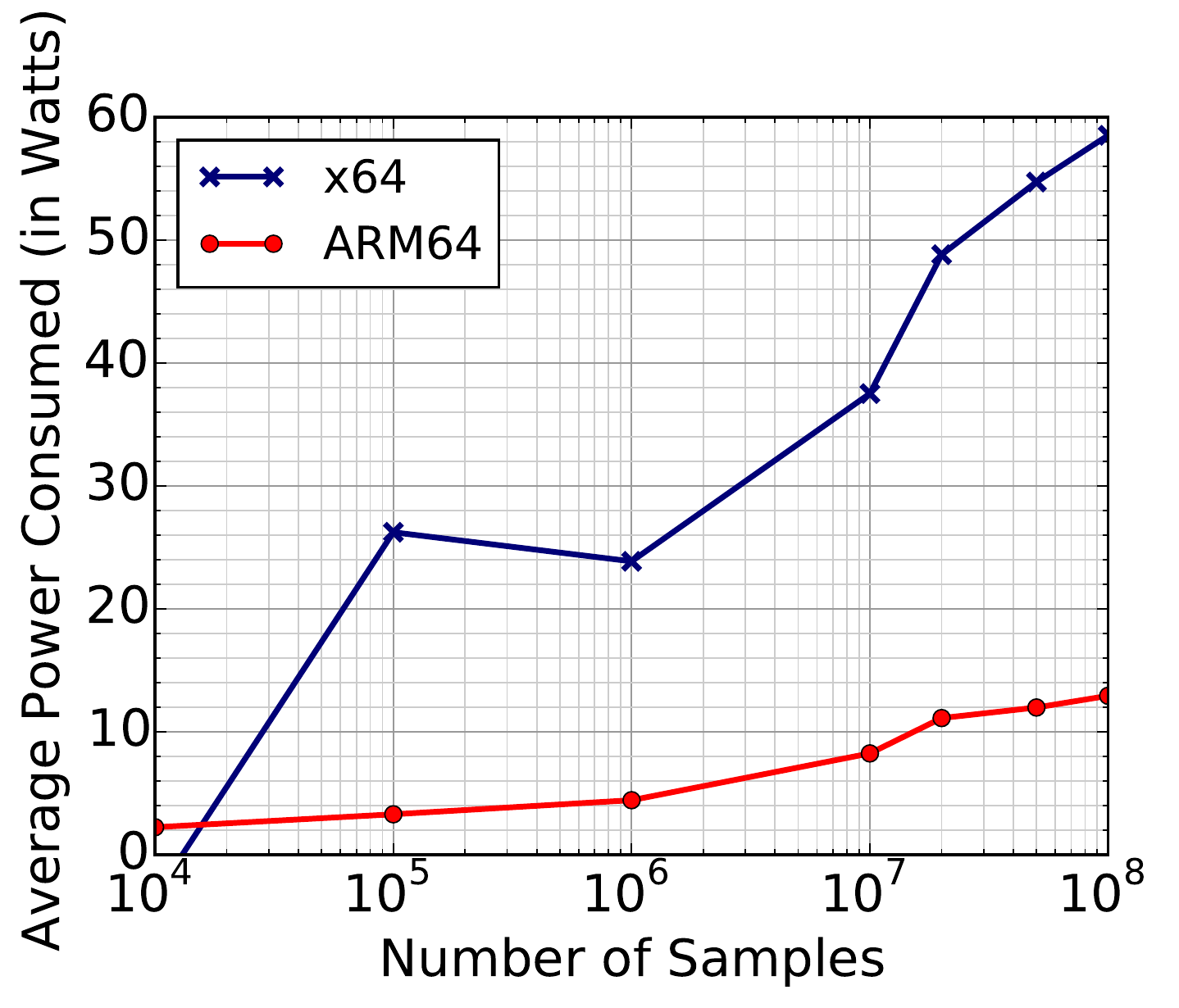}
		\label{fig:incpow:kmeans}
	}
	\caption{Average incremental power consumption above base load, for different data sizes}
\label{fig:incpow}
\end{figure}

Lastly, we examine the incremental power consumed by a benchmark as the workload size increases. Here, we subtract the base load from the average of the power loads sampled for a benchmark on a dataset, and plot the incremental average power consumed. For brevity, we only show plots for Nutch and K-Means in Figs.~\ref{fig:incpow:nutch} and \ref{fig:incpow:kmeans}, which behave differently. We see that Nutch has a slight negative slope for incremental power load as the data size increases in log scale on X Axis, while K-Means shows a clear positive slope with the data size. This trend is seen for both ARM64 and x64 servers. This again correlates strongly with the average CPU\% for these benchmarks and workloads. Nutch's CPU\% falls from an average $45-50\%$ for the smallest size to $20-25\%$ for the largest, for both servers, with median values flat at $20\%$. This is due to just a single Reducer task (1 core) being active for an increasingly longer time, as discussed before. K-Means, on the other hand, sees its CPU\% quickly grow from an average $36\%$ to $76\%$ between its smallest and largest sizes for both servers, with a similar increase in the median values as well. Thus, even with Big Data workloads, the energy costs are dominated by the processor utilization.

\section{Conclusion}
\label{sec:conclusion}
In this paper, we have presented results from evaluating an ARM64 server based on the recent AMD Opteron~A1100 SoC for Big Data workloads, and compared it with a similarly configured AMD Opteron~3380 x64 server. 
The results indicate that the ARM server shows comparable performance to the x64 server for integer-based workloads, and for smaller-sized floating-point workloads. For larger FP applications, its slower FPU impacts the performance. But here too, with tuning Hadoop to expose data parallelism, the ARM64 server can approach the performance of the x64 server, which is limited by having a faster FPU shared by pairs of cores.
An energy analysis shows the ARM64 server to have a $3\times$ smaller base power load than the x64 server, and a similar reduction in incremental load even when running the Big Data applications. We also see that the
EDP is better for ARM64 by up to $71\%$ compared to the x64 server.

At the high level, our experiments offers promise on leveraging ARM64 servers for Big Data applications within data centers. While our observations on ARM64's energy efficiency may appear intuitive, our work is the first to validate this empirically for Big Data workloads, and on commercially available ARM64 commodity servers. Somewhat less intuitive is our observation that the computational performance does not suffer significantly as a consequence. These are useful insights for data science application developers, Big Data platform architects, and IaaS and PaaS Cloud providers considering ARM in their data centers.

This study also opens up several interesting questions that require a detailed analysis, and motivate future work. We need to understand the disk performance, by examining the block sizes and other file-system factors, to help explain the ARM server's better I/O performance for smaller data sizes. 
It is necessary to look at hardware counters to ascertain what key architectural differences such as vector instructions play a role in the performance distinctions, beyond the observations on the FPUs and clock speeds -- neither processor offer an easy way to access their on-chip counters, which limited our current analysis. In their absence, we can also complement the workloads with more computational micro-benchmarks such as the HPL we attempted. It would also be compelling to compare ARM64 against a contemporary Intel Xeon server, pervasive in data centers, than just the AMD Opteron server that was readily available to us with a comparable configuration. 
It will also be interesting to examine the impact of containerization and virtualization on these two server platforms as they offer pathways for deployment in private and public Clouds. Lastly, exploring other Big Data workloads for stream processing and graph analytics is also planned.

\section*{Acknowledgments}
We thank Sarthak Sharma from the DREAM:Lab for setting up the initial Softiron server and benchmarking environment. We also thank Dr.T.V.Prabhakar and his staff at DESE, IISc for access to high-precision energy measurement instruments.

\bibliographystyle{IEEEtran}
\footnotesize{
\bibliography{arxiv2}
}

\end{document}